\documentstyle[epsfig,a4p,12pt,rotating]{article}
\begin{document}

%
%
\newcommand{\PPEnum}    {CERN-EP/98-122}
\newcommand{\PNnum}     {OPAL Physics Note PN-341}
\newcommand{\TNnum}     {OPAL Technical Note TN-xxx}
\newcommand{\Date}      {24$^{\mathrm{th}}$ July, 1998}
\newcommand{\Author}    {D.~I.~Futyan, K.~Roscoe, G.~W.~Wilson, T.~R.~Wyatt.}
\newcommand{\MailAddr}  {futyand@hpopb1.cern.ch}
\newcommand{\EdBoard}   {R.~Barlow, C.~Couyoumtzelis, I.~Fleck, S.~Talbot.}
\newcommand{\DraftVer}  {Version 2.0}
\newcommand{\DraftDate} {\today}
\newcommand{\TimeLimit} {Comments to wyatt@hepmail.ph.man.ac.uk by Wednesday, July~22$^{\mathrm{nd}}$ 1998, 12h00, please.}

\def\toprule{\noalign{\hrule \medskip}}
\def\midrule{\noalign{\medskip\hrule }}
\def\botrule{\noalign{\medskip\hrule }}
\setlength{\parskip}{\medskipamount}


\newcommand{\ee}{{\mathrm e}^+ {\mathrm e}^-}
\newcommand{\sq}{\tilde{\mathrm q}}
\newcommand{\seff}{\tilde{\mathrm f}}
\newcommand{\sele}{\tilde{\mathrm e}}
\newcommand{\sell}{\tilde{\ell}}
\newcommand{\snu}{\tilde{\nu}}
\newcommand{\smu}{\tilde{\mu}}
\newcommand{\stau}{\tilde{\tau}}
\newcommand{\chp}{\tilde{\chi}^+_1}
\newcommand{\chpm}{\tilde{\chi}^\pm_1}
\newcommand{\nt}{\tilde{\chi}^0}
\newcommand{\qq}{{\mathrm q}\bar{\mathrm q}}
\newcommand{\sleppair}{\sell^+ \sell^-}
\newcommand{\nunu}{\nu \bar{\nu}}
\newcommand{\mumu}{\mu^+ \mu^-}
\newcommand{\tautau}{\tau^+ \tau^-}
\newcommand{\ellell}{\ell^+ \ell^-}
\newcommand{\nulqq}{\nu \ell {\mathrm q} \bar{\mathrm q}'}
\newcommand{\MZ}{M_{\mathrm Z}}

\newcommand {\stopm}         {\tilde{\mathrm{t}}_{1}}
\newcommand {\stops}         {\tilde{\mathrm{t}}_{2}}
\newcommand {\stopbar}       {\bar{\tilde{\mathrm{t}}}_{1}}
\newcommand {\stopx}         {\tilde{\mathrm{t}}}
\newcommand {\sneutrino}     {\tilde{\nu}}
\newcommand {\slepton}       {\tilde{\ell}}
\newcommand {\stopl}         {\tilde{\mathrm{t}}_{\mathrm L}}
\newcommand {\stopr}         {\tilde{\mathrm{t}}_{\mathrm R}}
\newcommand {\stoppair}      {\tilde{\mathrm{t}}_{1}
\bar{\tilde{\mathrm{t}}}_{1}}
\newcommand {\gluino}        {\tilde{\mathrm g}}

\newcommand {\neutralino}    {\tilde{\chi }^{0}_{1}}
\newcommand {\neutrala}      {\tilde{\chi }^{0}_{2}}
\newcommand {\neutralb}      {\tilde{\chi }^{0}_{3}}
\newcommand {\neutralc}      {\tilde{\chi }^{0}_{4}}
\newcommand {\bino}          {\tilde{\mathrm B}^{0}}
\newcommand {\wino}          {\tilde{\mathrm W}^{0}}
\newcommand {\higginoa}      {\tilde{\rm H_{1}}^{0}}
\newcommand {\higginob}      {\tilde{\mathrm H_{1}}^{0}}
\newcommand {\chargino}      {\tilde{\chi }^{\pm}_{1}}
\newcommand {\charginop}     {\tilde{\chi }^{+}_{1}}
\newcommand {\KK}            {{\mathrm K}^{0}-\bar{\mathrm K}^{0}}
\newcommand {\ff}            {{\mathrm f} \bar{\mathrm f}}
\newcommand {\bstopm} {\mbox{$\boldmath {\tilde{\mathrm{t}}_{1}} $}}
\newcommand {\Mt}            {M_{\mathrm t}}
\newcommand {\mscalar}       {m_{0}}
\newcommand {\Mgaugino}      {M_{1/2}}
\newcommand {\rs}            {\sqrt{s}}
\newcommand {\WW}            {\mbox{${\mathrm W}^+{\mathrm W}^-$}}
\newcommand {\MGUT}          {M_{\mathrm {GUT}}}
\newcommand {\Zboson}        {${\mathrm Z}^{0}$}
\newcommand {\Wpm}           {{\mathrm W}^{\pm}}
\newcommand {\allqq}         {\sum_{q \neq t} q \bar{q}}
\newcommand {\mixang}        {\theta _{\mathrm {mix}}}
\newcommand {\thacop}        {\theta _{\mathrm {Acop}}}
\newcommand {\cosjet}        {\cos\thejet}
\newcommand {\costhr}        {\cos\thethr}
\newcommand {\djoin}         {d_{\mathrm{join}}}
\newcommand {\mstop}         {m_{\stopm}}
\newcommand {\msell}         {m_{\sell}}
\newcommand {\mchi}          {m_{\neutralino}}
\newcommand {\pp}{p \bar{p}}

\newcommand{\epair}{\mbox{${\mathrm e}^+{\mathrm e}^-$}}
\newcommand{\mupair}{\mbox{$\mu^+\mu^-$}}
\newcommand{\taupair}{\mbox{$\tau^+\tau^-$}}
\newcommand{\qpair}{\mbox{${\mathrm q}\overline{\mathrm q}$}}
\newcommand{\eeee}{\mbox{\epair\epair}}
\newcommand{\eemumu}{\mbox{\epair\mupair}}
\newcommand{\eetautau}{\mbox{\epair\taupair}}
\newcommand{\eeqq}{\mbox{\epair\qpair}}
\newcommand{\fs}{ final states}
\newcommand{\epairf}{\mbox{\epair\fs}}
\newcommand{\mupairf}{\mbox{\mupair\fs}}
\newcommand{\taupairf}{\mbox{\taupair\fs}}
\newcommand{\qpairf}{\mbox{\qpair\fs}}
\newcommand{\eeeef}{\mbox{\eeee\fs}}
\newcommand{\eemumuf}{\mbox{\eemumu\fs}}
\newcommand{\eetautauf}{\mbox{\eetautau\fs}}
\newcommand{\eeqqf}{\mbox{\eeqq\fs}}
\newcommand{\ffff}{four fermion final states}
\newcommand{\llnunu}{\mbox{$\ell^+\nu\,\ell^-\nbar$}}
\newcommand{\lnuqq}{\mbox{\lept\nubar\qpair}}
\newcommand{\zee}{\mbox{Zee}}
\newcommand{\zzg}{\mbox{ZZ/Z$\gamma$}}
\newcommand{\wenu}{\mbox{We$\nu$}}
\newcommand{\wwllnunu}{\mbox{$\wpair\rightarrow\llnunu$}}

\newcommand{\el}{\mbox{${\mathrm e}^-$}}
\newcommand{\selem}{\mbox{$\tilde{\mathrm e}^-$}}
\newcommand{\smum}{\mbox{$\tilde\mu^-$}}
\newcommand{\staum}{\mbox{$\tilde\tau^-$}}
\newcommand{\slept}{\mbox{$\tilde{\ell}^\pm$}}
\newcommand{\sleptm}{\mbox{$\tilde{\ell}^-$}}
\newcommand{\lept}{\mbox{$\ell^-$}}
\newcommand{\Hl}{\mbox{$\mathrm{L}^\pm$}}
\newcommand{\Hm}{\mbox{$\mathrm{L}^-$}}
\newcommand{\Hnu}{\mbox{$\nu_{\mathrm{L}}$}}
\newcommand{\nul}{\mbox{$\nu_\ell$}}
\newcommand{\nubar}{\mbox{$\overline{\nu}_\ell$}}
\newcommand{\nbar}{\mbox{$\overline{\nu}$}}
\newcommand{\spair}{\mbox{$\tilde{\ell}^+\tilde{\ell}^-$}}
\newcommand{\lpair}{\mbox{$\ell^+\ell^-$}}
\newcommand{\staupair}{\mbox{$\tilde{\tau}^+\tilde{\tau}^-$}}
\newcommand{\smupair}{\mbox{$\tilde{\mu}^+\tilde{\mu}^-$}}
\newcommand{\selepair}{\mbox{$\tilde{\mathrm e}^+\tilde{\mathrm e}^-$}}
\newcommand{\ch}{\mbox{$\tilde{\chi}^\pm_1$}}
\newcommand{\chpair}{\mbox{$\tilde{\chi}^+_1\tilde{\chi}^-_1$}}
\newcommand{\chm}{\mbox{$\tilde{\chi}^-_1$}}
\newcommand{\chmp}{\mbox{$\tilde{\chi}^\pm_1$}}
\newcommand{\chz}{\mbox{$\tilde{\chi}^0_1$}}
\newcommand{\dch}{\mbox{\chm$\rightarrow$\lept\chz\nubar}}
\newcommand{\dslept}{\mbox{\sleptm$\rightarrow$\lept\chz}}
\newcommand{\dH}{\mbox{H$^{+}\rightarrow\tau^{+}\nu_{\tau}$}}
\newcommand{\dHL}{\mbox{\Hm$\rightarrow\lept\nubar\Hnu$}}
\newcommand{\mch}{\mbox{$m_{\tilde{\chi}^\pm_1}$}}
\newcommand{\mslept}{\mbox{$m_{\tilde{\ell}}$}}
\newcommand{\mstau}{\mbox{$m_{\tilde\tau^-}$}}
\newcommand{\msmu}{\mbox{$m_{\tilde\mu^-}$}}
\newcommand{\msele}{\mbox{$m_{\tilde{\mathrm e}^-}$}}
\newcommand{\mchz}{\mbox{$m_{\tilde{\chi}^0_1}$}}
\newcommand{\dm}{\mbox{$\Delta m$}}
\newcommand{\dmch}{\mbox{$\Delta m_{\ch-\chz}$}}
\newcommand{\dmslept}{\mbox{$\Delta m_{\slept-\chz}$}}
\newcommand{\dmhl}{\mbox{$\Delta m_{\Hl-\Hnu}$}}
\newcommand{\w}{\mbox{W$^\pm$}}
\newcommand{\mH}{\mbox{$m_{\mathrm{H}^+}$}}
\newcommand{\chargthree}{\chpair (3-body decays)}
\newcommand{\chargtwo}{\chpair (2-body decays)}
\newcommand{\mf}     {\mbox{$(m-15)/2$}}
\newcommand{\dchtwo}{\mbox{$\chpm \rightarrow \ell^\pm {\tilde{\nu}_\ell}$}}
\newcommand{\dchthree}{\mbox{$\chmp\rightarrow{\mathrm W}^\pm\chz\rightarrow\ell^\pm\nu\chz$}}
\newcommand{\msnu}{\mbox{$m_{\snu}$}}
\newcommand{\hpair}{\mbox{$\mathrm{H}^+\mathrm{H}^-$}}
\newcommand{\chargthreee}{\chpair (3-body decays: \dchthree )}
\newcommand{\chargtwoo}{\chpair (2-body decays: \dchtwo )}
\newcommand{\Nninef}   {\mbox{$N_{95}(N,\mu _B)$}}
\newcommand{\Nnine}   {\mbox{$N_{95}$}}

\newcommand{\acopc}{\mbox{$\phi^{\mathrm{acop}}$}}
\newcommand{\acolc}{\mbox{$\theta^{\mathrm{acol}}$}}
\newcommand{\acop}{\mbox{$\phi_{\mathrm{acop}}$}}
\newcommand{\acol}{\mbox{$\theta_{\mathrm{acol}}$}}
\newcommand{\pz}{\mbox{$p_{\mathrm{z}}^{\mathrm{miss}}$}}
\newcommand{\ptevt}{\mbox{$p_{t}^{\mathrm{miss}}$}}
\newcommand{\ptaxic}{\mbox{$a_{t}^{\mathrm{miss}}$}}
\newcommand{\stevt}{\mbox{$p_{t}^{\mathrm{miss}}$/\Ebeam}}
\newcommand{\staxic}{\mbox{$a_{t}^{\mathrm{miss}}$/\Ebeam}}
\newcommand{\dptaxic}{\mbox{missing $p_{t}$ wrt. event axis \ptaxic}}
\newcommand{\cosevt}{\mbox{$\mid\cos\theta_{\mathrm{p}}^{\mathrm{miss}}\mid$}}
\newcommand{\axicos}{\mbox{$\mid\cos\theta_{\mathrm{a}}^{\mathrm{miss}}\mid$}}
\newcommand{\pthet}{\mbox{$\theta_{\mathrm{p}}^{\mathrm{miss}}$}}
\newcommand{\athet}{\mbox{$\theta_{\mathrm{a}}^{\mathrm{miss}}$}}
\newcommand{\dcosevt}{\mbox{$\mid\cos\theta\mid$ of missing p$_{t}$}}
\newcommand{\daxicos}{\mbox{$\mid\cos\theta\mid$ of missing p$_{t}$ wrt. event
axis}}
\newcommand{\efdsw}{\mbox{$x_{\mathrm{FDSW}}$}}
\newcommand{\acopf}{\mbox{$\Delta\phi_{\mathrm{FDSW}}$}}
\newcommand{\acopm}{\mbox{$\Delta\phi_{\mathrm{MUON}}$}}
\newcommand{\acopt}{\mbox{$\Delta\phi_{\mathrm{trk}}$}}
\newcommand{\po}{\mbox{$E_{\mathrm{isol}}^\gamma$}}
\newcommand{\qprod}{\mbox{$q1$$*$$q2$}}
\newcommand{\lcode}{lepton identification code}
\newcommand{\nctro}{\mbox{$N_{\mathrm{trk}}^{\mathrm{out}}$}}
\newcommand{\necao}{\mbox{$N_{\mathrm{ecal}}^{\mathrm{out}}$}}
\newcommand{\mout}{\mbox{$m^{\mathrm{out}}$}}
\newcommand{\nctec}{\mbox{\nctro$+$\necao}}
\newcommand{\gfract}{\mbox{$F_{\mathrm{good}}$}}
\newcommand{\zz}       {\mbox{$|z_0|$}}
\newcommand{\dz}       {\mbox{$|d_0|$}}
\newcommand{\sint}      {\mbox{$\sin\theta$}}
\newcommand{\cost}      {\mbox{$\cos\theta$}}
\newcommand{\mcost}     {\mbox{$|\cos\theta|$}}
\newcommand{\dedx}     {\mbox{$dE/dx$}}
\newcommand{\wdedx}     {\mbox{$W_{dE/dx}$}}
\newcommand{\xe}     {\mbox{$x_E$}}

\newcommand{\p}     {\mbox{$\pm$}}
\newcommand{\ssix}     {\mbox{$\protect\sqrt{s}$~=~161~GeV}}
\newcommand{\sseven}     {\mbox{$\protect\sqrt{s}$~=~172~GeV}}
\newcommand{\seight}     {\mbox{$\protect\sqrt{s}$~=~183~GeV}}
\newcommand{\sthree}     {\mbox{$\protect\sqrt{s}$~=~130--136~GeV}}
\newcommand{\mrecoil}     {\mbox{$m_{\mathrm{recoil}}$}}
\newcommand{\llmass}     {\mbox{$m_{ll}$}}
\newcommand{\sml}{\mbox{Standard Model \llnunu\ events}}
\newcommand{\sme}{\mbox{Standard Model events}}
\newcommand{\sig}{events containing a lepton pair plus missing transverse momentum}
\newcommand{\wpair}{\mbox{$\mathrm{W}^+\mathrm{W}^-$}}
\newcommand{\dW}{\mbox{W$^-\rightarrow\lept\nubar$}}
\newcommand{\dsele}{\mbox{$\selem\rightarrow\mathrm{e}^-\chz$}}
\newcommand{\dstau}{\mbox{$\staum\rightarrow\tau^-\chz$}}
\newcommand{\eeeell}{\mbox{\epair$\rightarrow$\epair\lpair}}
\newcommand{\eell}{\mbox{\epair\lpair}}
\newcommand{\llgam}{\mbox{$\ell^+\ell^-(\gamma)$}}
\newcommand{\nunugam}{\mbox{$\nu\bar{\nu}\gamma\gamma$}}
\newcommand{\acope}{\mbox{$\Delta\phi_{\mathrm{EE}}$}}
\newcommand{\nee}{\mbox{N$_{\mathrm{EE}}$}}
\newcommand{\eesum}{\mbox{$\Sigma_{\mathrm{EE}}$}}
\newcommand{\at}{\mbox{$a_{t}$}}
\newcommand{\spp}{\mbox{$p$/\Ebeam}}
\newcommand{\acoph}{\mbox{$\Delta\phi_{\mathrm{HCAL}}$}}
\newcommand{\ACOP}{\mbox{$\phi_{\mathrm{acop}}$}}
\newcommand{\XT}{\mbox{$x_T$}}
\newcommand{\XONE}{\mbox{$x_1$}}
\newcommand{\XTWO}{\mbox{$x_2$}}
\newcommand{\MLL}{\mbox{$m_{\ell\ell}$}}
\newcommand{\MRECOIL}{\mbox{$m_{\mathrm{recoil}}$}}
\newcommand {\mm}       {\mu^+ \mu^-}
\newcommand {\emu}         {\mathrm{e}^{\pm} \mu^{\mp}}
\newcommand {\et}         {\mathrm{e}^{\pm} \tau^{\mp}}
\newcommand {\mt}         {\mu^{\pm} \tau^{\mp}}
\newcommand {\lemu}       {\ell=\mathrm{e},\mu}
\newcommand{\Zz}{\mbox{${\mathrm{Z}^0}$}}

\newcommand{\ZP}[3]    {Z. Phys. {\bf C#1} (#2) #3.}
\newcommand{\PL}[3]    {Phys. Lett. {\bf B#1} (#2) #3.}
\newcommand{\etal}     {{\it et al}.}

\newcommand{\Ecm}{\mbox{$E_{\mathrm{cm}}$}}
\newcommand{\Ebeam}{\mbox{$E_{\mathrm{beam}}$}}
\newcommand{\ipb}{\mbox{pb$^{-1}$}}
\newcommand{\wrt}{with respect to}
\newcommand{\sm}{Standard Model}
\newcommand{\smb}{Standard Model background}
\newcommand{\smp}{Standard Model processes}
\newcommand{\smc}{Standard Model Monte Carlo}
\newcommand{\mc}{Monte Carlo}
\newcommand{\btb}{back-to-back}
\newcommand{\tp}{two-photon}
\newcommand{\tpb}{two-photon background}
\newcommand{\tpp}{two-photon processes}
\newcommand{\lp}{lepton pairs}
\newcommand{\vto}{\mbox{$\tau$ veto}}

\newcommand{\gsim}{\;\raisebox{-0.9ex}
           {$\textstyle\stackrel{\textstyle >}{\sim}$}\;}
\newcommand{\lsim}{\;\raisebox{-0.9ex}{$\textstyle\stackrel{\textstyle<}
           {\sim}$}\;}

\newcommand{\degree}    {^\circ}
%
\newcommand{\roots}     {\protect\sqrt{s}}
%
%
\newcommand{\thrust}    {T}
\newcommand{\nthrust}   {\hat{n}_{\mathrm{thrust}}}
\newcommand{\thethr}    {\theta_{\,\mathrm{thrust}}}
\newcommand{\phithr}    {\phi_{\mathrm{thrust}}}
\newcommand{\acosthr}   {|\cos\thethr|}
\newcommand{\thejet}    {\theta_{\,\mathrm{jet}}}
\newcommand{\acosjet}   {|\cos\thejet|}
\newcommand{\thmiss}    { \theta_{miss} }
\newcommand{\cosmiss}   {| \cos \thmiss |}
%
%
\newcommand{\Evis}      {E_{\mathrm{vis}}}
\newcommand{\Rvis}      {E_{\mathrm{vis}}\,/\roots}
\newcommand{\Mvis}      {M_{\mathrm{vis}}}
\newcommand{\Rbal}      {R_{\mathrm{bal}}}
%
%
\newcommand{\phiacop}   {\phi_{\mathrm{acop}}}
%
%
\newcommand{\LS}      {L_{S}}
\newcommand{\LB}      {L_{B}}
\newcommand{\LR}      {L_{R}}
\newcommand{\PS}      {\mbox{$P(x_i,S)$}}
\newcommand{\PB}      {\mbox{$P(x_i,B)$}}
\newcommand{\signine}   {\mbox{$\sigma_{95}$}}
\newcommand{\expsig}   {\mbox{$\langle\signine\rangle$}}
%
%
%
\newcommand{\PhysLett}  {Phys.~Lett.}
\newcommand{\PRL} {Phys.~Rev.\ Lett.}
\newcommand{\PhysRep}   {Phys.~Rep.}
\newcommand{\PhysRev}   {Phys.~Rev.}
\newcommand{\NPhys}  {Nucl.~Phys.}
\newcommand{\NIM} {Nucl.~Instr.\ Meth.}
\newcommand{\CPC} {Comp.~Phys.\ Comm.}
\newcommand{\ZPhys}  {Z.~Phys.}
\newcommand{\IEEENS} {IEEE Trans.\ Nucl.~Sci.}
%
%
\newcommand{\OPALColl}  {OPAL Collaboration}
%
\newcommand{\onecol}[2] {\multicolumn{1}{#1}{#2}}
\newcommand{\ra}        {\rightarrow}   


\begin{titlepage}
%
%
\begin{center}
    \large
    EUROPEAN LABORATORY FOR PARTICLE PHYSICS
\end{center}
\begin{flushright}
    \large
    \PPEnum\\
    \Date
\end{flushright}

%
%
\begin{center}
    \huge\bf\boldmath
    Search for Acoplanar Lepton Pair Events in e$^+$e$^-$ Collisions\\
    at $\sqrt{s} = 161$, 172 and 183 GeV
\end{center}
%
%
\begin{center}
 \Large
 The OPAL Collaboration \\
\bigskip
\bigskip
\bigskip
\end{center}
%
%
\begin{abstract}

A selection of di-lepton events with significant missing transverse momentum 
has been performed using a total data sample of 77.0~pb$^{-1}$
at e$^+$e$^-$ centre-of-mass energies 
of 161~GeV, 172~GeV and 183~GeV.
The observed numbers of events: four at 161 GeV, nine at 172 GeV, and
78 at 183~GeV,
are consistent with the numbers expected from Standard Model processes, 
which arise predominantly from \wpair\ production with each W decaying
leptonically. 
This topology is an experimental signature
also for the pair production of new particles that decay
to a charged lepton accompanied by one or more
invisible particles.
Further event selection criteria are described that optimise 
the sensitivity to particular new physics channels.
No evidence for new phenomena is apparent and model independent limits 
on the production cross-section times branching ratio squared
for various new physics processes are presented.
Assuming a 100\% branching ratio for the decay
$\sell^\pm_R \rightarrow  {\ell^\pm} \nt_1$, we exclude at 95\% CL:
right-handed smuons with masses below 65~GeV for 
\mbox{$\msmu - \mchz > 2$}~GeV and
right-handed staus with masses below 64~GeV for 
\mbox{$\mstau - \mchz > 10$}~GeV.
Right-handed selectrons are excluded at 95\% CL for 
masses below 77~GeV for \mbox{$\msele - \mchz > 5$}~GeV
within the framework of the
Minimal Supersymmetric Standard Model assuming
$\mu < -100$~GeV and $\tan{\beta} = 1.5$.

\end{abstract}
\bigskip
\bigskip
\begin{center}
{\large (Submitted to Eur.~Phys.~J.~C.)}
\end{center}

\end{titlepage}
 
\begin{center}{
G.\thinspace Abbiendi$^{  2}$,
K.\thinspace Ackerstaff$^{  8}$,
G.\thinspace Alexander$^{ 23}$,
J.\thinspace Allison$^{ 16}$,
N.\thinspace Altekamp$^{  5}$,
K.J.\thinspace Anderson$^{  9}$,
S.\thinspace Anderson$^{ 12}$,
S.\thinspace Arcelli$^{ 17}$,
S.\thinspace Asai$^{ 24}$,
S.F.\thinspace Ashby$^{  1}$
% ...
}\end{center}\bigskip
\bigskip
$^{  1}$School of Physics and Astronomy, University of Birmingham,
Birmingham B15 2TT, UK
\newline
$^{  2}$Dipartimento di Fisica dell' Universit\`a di Bologna and INFN,
I-40126 Bologna, Italy
\newline
$^{  3}$Physikalisches Institut, Universit\"at Bonn,
D-53115 Bonn, Germany
\newline
$^{  4}$Department of Physics, University of California,
Riverside CA 92521, USA
\newline
$^{  5}$Cavendish Laboratory, Cambridge CB3 0HE, UK
\newline
$^{  6}$Ottawa-Carleton Institute for Physics,
Department of Physics, Carleton University,
Ottawa, Ontario K1S 5B6, Canada
\newline
$^{  7}$Centre for Research in Particle Physics,
Carleton University, Ottawa, Ontario K1S 5B6, Canada
\newline
$^{  8}$CERN, European Organisation for Particle Physics,
CH-1211 Geneva 23, Switzerland
\newline
$^{  9}$Enrico Fermi Institute and Department of Physics,
University of Chicago, Chicago IL 60637, USA
\newline
$^{ 10}$Fakult\"at f\"ur Physik, Albert Ludwigs Universit\"at,
D-79104 Freiburg, Germany
\newline
$^{ 11}$Physikalisches Institut, Universit\"at
Heidelberg, D-69120 Heidelberg, Germany
\newline
$^{ 12}$Indiana University, Department of Physics,
Swain Hall West 117, Bloomington IN 47405, USA
\newline
$^{ 13}$Queen Mary and Westfield College, University of London,
London E1 4NS, UK
\newline
$^{ 14}$Technische Hochschule Aachen, III Physikalisches Institut,
Sommerfeldstrasse 26-28, D-52056 Aachen, Germany
\newline
$^{ 15}$University College London, London WC1E 6BT, UK
\newline
$^{ 16}$Department of Physics, Schuster Laboratory, The University,
Manchester M13 9PL, UK
\newline
$^{ 17}$Department of Physics, University of Maryland,
College Park, MD 20742, USA
\newline
$^{ 18}$Laboratoire de Physique Nucl\'eaire, Universit\'e de Montr\'eal,
Montr\'eal, Quebec H3C 3J7, Canada
\newline
$^{ 19}$University of Oregon, Department of Physics, Eugene
OR 97403, USA
\newline
$^{ 20}$CLRC Rutherford Appleton Laboratory, Chilton,
Didcot, Oxfordshire OX11 0QX, UK
\newline
$^{ 22}$Department of Physics, Technion-Israel Institute of
Technology, Haifa 32000, Israel
\newline
$^{ 23}$Department of Physics and Astronomy, Tel Aviv University,
Tel Aviv 69978, Israel
\newline
$^{ 24}$International Centre for Elementary Particle Physics and
Department of Physics, University of Tokyo, Tokyo 113, and
Kobe University, Kobe 657, Japan
\newline
$^{ 25}$Institute of Physical and Environmental Sciences,
Brunel University, Uxbridge, Middlesex UB8 3PH, UK
\newline
$^{ 26}$Particle Physics Department, Weizmann Institute of Science,
Rehovot 76100, Israel
\newline
$^{ 27}$Universit\"at Hamburg/DESY, II Institut f\"ur Experimental
Physik, Notkestrasse 85, D-22607 Hamburg, Germany
\newline
$^{ 28}$University of Victoria, Department of Physics, P O Box 3055,
Victoria BC V8W 3P6, Canada
\newline
$^{ 29}$University of British Columbia, Department of Physics,
Vancouver BC V6T 1Z1, Canada
\newline
$^{ 30}$University of Alberta,  Department of Physics,
Edmonton AB T6G 2J1, Canada
\newline
$^{ 31}$Duke University, Dept of Physics,
Durham, NC 27708-0305, USA
\newline
$^{ 32}$Research Institute for Particle and Nuclear Physics,
H-1525 Budapest, P O  Box 49, Hungary
\newline
$^{ 33}$Institute of Nuclear Research,
H-4001 Debrecen, P O  Box 51, Hungary
\newline
$^{ 34}$Ludwigs-Maximilians-Universit\"at M\"unchen,
Sektion Physik, Am Coulombwall 1, D-85748 Garching, Germany
\newline
\bigskip\newline
$^{  a}$ and at TRIUMF, Vancouver, Canada V6T 2A3
\newline
$^{  b}$ and Royal Society University Research Fellow
\newline
$^{  c}$ and Institute of Nuclear Research, Debrecen, Hungary
\newline
$^{  d}$ and Department of Experimental Physics, Lajos Kossuth
University, Debrecen, Hungary
\newline
$^{  e}$ on leave of absence from the University of Freiburg
\newline
\clearpage

\section{Introduction}
\label{sec:intro}

We report on the selection of events containing two charged leptons and 
significant missing transverse momentum.
Data are analysed at e$^+$e$^-$ centre-of-mass energies of 161, 172
and 183\footnote{
In the 1997 LEP run most of the data were collected at  
e$^+$e$^-$ centre-of-mass energies of between 181.8 and 183.8~GeV.
The luminosity weighted average  centre-of-mass energy was 182.7~GeV.
}~GeV
with integrated luminosities corresponding to
 10.3 \ipb , 10.3 \ipb\ and 56.4 \ipb , respectively.
The number and properties of the observed events are found to be
consistent with the expectations for  \smp , which are dominated by 
the \llnunu\ final state arising from \wpair\ production in which both W's
decay leptonically: $\dW$.

In most respects the analysis method follows closely our previously
published search at 161 and 172 GeV~\cite{ppe124}.  
The event selection is performed in two stages.
The first stage consists of a general selection for all the possible
\sig\ (section~\ref{sec:gen}).
In this context the \sm\ \llnunu\ events are considered as signal
in addition to the possible new physics sources.
\smp\ that do not lead to \llnunu\ final states --- e.g., \eell\ and
\llgam\ --- are considered as background and are reduced to a
rather low level.
In the second stage selection the detailed properties of
the events are used to separate as far as  possible the events
consistent with
potential new physics sources from  \wpair\ and other \smp\
  (section~\ref{sec:addkin}).

In companion papers~\cite{Wmass,wwpaper172,wwpaper} we use the selected
event samples to measure the production of  \wwllnunu\ events.
We present here the results of searches for several anomalous sources of 
such events.
We consider the pair production of new particles that decay
to produce a charged lepton  accompanied by one or more
invisible particles, such as neutrinos or 
the hypothesised lightest stable supersymmetric~\cite{SUSY} particle (LSP), 
which may be the lightest neutralino, $\nt_1$, or the gravitino.
Specifically, we consider the following new particle decays:
\begin{description}
\item[charged scalar leptons (sleptons):]
$\sell^\pm \rightarrow  {\ell^\pm} \nt_1$,
where $\sell^\pm$ may be a selectron ($\sele$), smuon ($\smu$) or stau
($\stau$) and $\ell^\pm$ is the corresponding charged lepton.
\item[charged Higgs bosons:] $\mathrm{H}^{\pm} \rightarrow \tau^\pm \nu_\tau$.
\item[charginos:] $\chpm \rightarrow \ell^\pm \snu$ (``2-body'' decays)
\ \   or \ \ 
$\chpm \rightarrow \ell^\pm \nu \chz$ (``3-body'' decays).
\end{description}
Searches for sleptons at LEP2 using this topology have been presented
also by other collaborations~\cite{olep}. 

In this paper we describe fully only those respects in which the
analysis differs significantly from~\cite{ppe124}.
These are:
\begin{itemize}
\item
Use of a new subdetector (the MIP plug~\cite{tenim}) 
to reduce background from four-fermion processes in which 
a minimum ionizing particle would otherwise have escaped detection in
the angular region\footnote{
A right-handed coordinate system is adopted,
in which the $x$-axis points to the centre of the LEP ring,
and positive $z$ is along  the electron beam direction.
The angles $\theta$ and $\phi$ are the polar and azimuthal angles,
respectively.} 
  $60 < \theta (\mathrm{mrad}) < 160$.
\item
The second stage event selection for smuons, staus and charged Higgs bosons
makes use of the charge-signed angular distribution of
the observed lepton candidates.
\item
Use of a likelihood technique in the second stage of event selection,
to determine whether an event is more
consistent with the signal or with the background hypothesis, and
applying cuts based on this information.
\item
In section~\ref{results}, 
the 95\% CL upper limits on  new particle production at \seight\ are
obtained by combining the data at the three centre-of-mass 
energies 161, 172 and 183~GeV using
the Likelihood Ratio method~\cite{LR}.
\end{itemize}
The MIP plug is available only for the data at 183~GeV.
The remaining modifications have been applied to the data at all
centre-of-mass energies.

\section{OPAL Detector and Monte Carlo Simulation}
\label{opaldet}

A detailed description of the  OPAL detector can be found 
elsewhere~\cite{OPAL-detector}.

The central detector consists of
a system of tracking chambers
providing charged particle tracking
over 96\% of the full solid 
angle
inside a 0.435~T uniform magnetic field parallel to the beam axis. 
It consists of a two-layer
silicon micro-strip vertex detector, a high precision drift chamber,
a large volume jet chamber and a set of $z$ chambers  that measure 
the track coordinates along the beam direction. 

A lead-glass electromagnetic
calorimeter located outside the magnet coil
covers the full azimuthal range with excellent hermeticity
in the polar angle range of $|\cos \theta |<0.82$ for the barrel
region and $0.81<|\cos \theta |<0.984$ for the endcap region (EE).
Electromagnetic calorimeters close to the beam axis 
complete the geometrical acceptance down to approximately 25 mrad.
These include 
the forward detectors (FD) which are
lead-scintillator sandwich calorimeters and, at smaller angles,
silicon tungsten calorimeters (SW)
located on both sides of the interaction point.
The gap between the EE and FD calorimeters
is instrumented with an additional lead-scintillator 
electromagnetic calorimeter,
called the gamma-catcher (GC).

The magnet return yoke is instrumented for hadron calorimetry 
and consists of barrel and endcap sections along with pole tip detectors that
together cover the region $|\cos \theta |<0.99$.
Outside the hadron calorimeter, four layers of muon chambers 
cover the polar angle range of $|\cos \theta |<0.98$. 
Arrays of thin scintillating tiles 
with embedded wavelength shifting 
fibre readout have been installed in the
endcap region to improve trigger performance, time resolution and hermeticity
for experiments at LEP~II~\cite{tenim}. 
Of particular relevance to this analysis are the four layers
of scintillating tiles (the MIP plug)  installed at 
each end of OPAL covering
the angular range $43 < \theta (\mathrm{mrad}) < 220$.

The following \smp\ are simulated at \seight .
4-fermion production is generated using grc4f~\cite{grc4f},
{\sc Pythia}~\cite{pythia} and {\sc Excalibur}~\cite{excalibur}.
Two-photon processes are generated using the program of
Vermaseren~\cite{vermaseren} and grc4f
for \eell , and 
using  {\sc Phojet}~\cite{phojet}, {\sc Herwig}~\cite{herwig}
  and grc4f for \eeqq .
Because of the large total cross-section for  \eeee , \eemumu\ and
\eeqq , soft cuts are applied at the generator level to preselect
events that might possibly lead to background in the selection of
\llnunu\ final states.
No generator level cuts are applied to the \eetautau\ generation.
The production of lepton pairs is generated using 
{\sc Bhwide}~\cite{bhwide} and {\sc Teegg}~\cite{teegg} for $\ee(\gamma)$, 
 and using {\sc
  Koralz}~\cite{koralz} for $\mumu(\gamma)$, $\tautau(\gamma)$ and
\nunugam .
The production of quark pairs, \qpair(g), is generated using {\sc Pythia}. 
The effective \mc\ integrated luminosities exceed that of the data by
factors that are typically of the order of one thousand and in all
cases are at least twenty.

Slepton pair production is generated using 
{\sc Susygen}~\cite{SUSYGEN}.
Charged Higgs boson pair production  is generated using
{\sc Hzha}~\cite{HZHA} and {\sc Pythia}.
Chargino pair production is generated using  {\sc Dfgt}~\cite{DFGT} and 
{\sc Susygen}.

All \sm\ and new physics \mc\ samples are processed with a full simulation  
of the OPAL detector \cite{gopal} and subjected to the same
reconstruction and analysis programs as used for the OPAL data.

\section{General Selection of Di-lepton Events with Significant Missing Momentum}
\label{sec:gen}

The general selection of di-lepton events with significant missing
 momentum is largely unchanged \wrt~\cite{ppe124}.
In selecting candidate events the missing
momentum is required to have a significant component 
in the plane perpendicular to the beam axis (\ptevt ).
This generally leads to an acoplanar\footnote{
The acoplanarity angle is defined  as 180$^{\circ}$ minus the angle
between the two lepton candidates in the plane transverse to the 
beam direction.} 
event topology.
Standard Model background with high energy particles
escaping down the beam pipe and giving rise 
to missing momentum along the beam axis is thus rejected.
A potential background arises from lepton
pairs  produced in two-photon processes
in which one of the initial state electrons  is scattered at a
significant angle to the beam direction. 
Such processes are suppressed by vetoing on energy being 
present in the forward region (SW, FD or GC calorimeters).

The OPAL detector provides hermetic coverage for electrons and photons
for $\theta > 25$~mrad.
However, prior to the data collection at \seight\ it was possible for a muon in
the angular range $60 < \theta (\mathrm{mrad}) < 160$ to escape
detection.
This led to a background to the general selection from \eemumu\ events
in which one electron and one muon were observed in the detector. 
In order to improve the detection of muons in the forward region, four layers
of scintillating tile detectors (the MIP plug) were installed at 
each end of OPAL.
In the data at \seight\ the OPAL detector provides hermetic 
coverage for muons for $\theta > 25$~mrad.

A new selection cut is introduced to make use of the  MIP plug detector.
Candidate events are vetoed if they contain
coincident hits in two or more scintillator layers in the MIP plug
 at the same $\phi$ and at the same end of OPAL, satisfying cuts on
 pulse height and timing.
The efficiency 
to detect a muon within the geometrical acceptance of the MIP plug
with these cuts is measured to be $80\pm 4\%$ by 
using \eemumu\ 
events in which a minimum ionizing track is observed in SW
within the MIP plug geometrical acceptance.
Note that this efficiency includes the effect of periods when the MIP
plug was not fully operational.
The cut on MIP plug activity
exclusively rejects three events in the data at \seight ;
this may be
compared with the \sm\ expectation of 2.5 events, of which 2.3 arise
from the \eemumu\ final state.
In addition to the cut on MIP plug activity, minor
changes to improve the rejection of \tpp\ and ``junk'' events 
arising from, e.g., beam-gas interactions have been made \wrt\ the
selection cuts described in~\cite{ppe124}. 

The numbers of events passing the general selection at each centre-of-mass
energy in the data are compared to the \smc\ predictions 
in table~\ref{tab-samples}. 
The total number of events predicted by the  \sm\ is given, 
together with a breakdown into the
contributions from individual processes.
The number of observed candidates  is consistent with the 
expectation from \sm\ sources, which is dominated by 
the \llnunu\ final state arising from \wpair\ production in which both W's
decay leptonically: $\dW$.
\begin{table}
\centering
\begin{tabular}{||r||r|r||r|r|r|r||}
\hline \hline
$\sqrt{s}$
 (GeV)& data & SM        & \llnunu   & \eell    & \eeqq    & \llgam\   \\
\hline \hline
  161 &    4 &  4.5\p0.3 &  3.4\p0.1 & 0.9\p0.2 & 0.1\p0.1 & 0.1\p0.0   \\
  172 &    9 & 11.6\p0.3 & 10.7\p0.1 & 0.8\p0.2 & 0.0\p0.0 & 0.1\p0.0   \\
  183 &   78 & 81.3\p0.7 & 77.6\p0.7 & 3.2\p0.0 & 0.0\p0.0 & 0.5\p0.2   \\
\hline \hline
\end{tabular}
\caption[]{\sl
  \protect{\parbox[t]{15cm}{
Comparison between data and \mc\ of the 
number of events passing the general selection at each centre-of-mass
energy. 
The total number of events predicted by the  \sm\ is given, 
together with a breakdown into the
contributions from individual processes.
At each centre-of-mass energy
the \sm\ \mc s are normalised to an integrated luminosity that
 corresponds to the collected experimental luminosity.
The \mc\ statistical errors are given. 
\label{tab-samples}
}} }
\end{table}

The second stage event selection to distinguish between 
\sm\ and new physics sources of lepton pair
events with missing momentum is described in section~\ref{sec:addkin}.
Discrimination is provided by information on the 
lepton identification, and the energy and $-q \cos\theta$ of
the observed lepton candidates, where
$q$ is the lepton charge. 
We check here on the degree to which these quantities are described by
the \smc .
The lepton identification information in the event sample produced
 by the general selection at \seight\  is
compared with the \smc\ in table~\ref{tab-dlept}.
\begin{table}
\centering
\begin{tabular}{||c|r|r||}
\hline\hline
Lepton identification                 &  data  &  SM   \\
\hline\hline
\epair\                               &  14    & 12.3  \\
\mupair\                              &  13    & 13.6  \\
$h^{\pm} h^{\mp}$                     &   1    &  2.3  \\
$\emu$                                &  20    & 25.6  \\
$\mathrm{e}^{\pm} h^{\mp}$            &   8    &  9.8  \\
$\mu^{\pm} h^{\mp}$                   &   8    &  9.3  \\
$\mathrm{e}^{\pm}$, unidentified      &   5    &  3.8  \\
$\mu^{\pm}$, unidentified             &   7    &  3.7  \\
$h^{\pm}$, unidentified               &   2    &  0.9  \\
\hline\hline
\end{tabular}
\caption[]{\sl
  \protect{\parbox[t]{15cm}{
The lepton identification information in the 
events passing the general selection
compared with the \smc\ at \seight .
  ``$h$'' means
that the lepton is identified neither as an electron nor muon and
so is probably the product of a hadronic tau decay.
Leptonic decays of taus are usually classified
as electron or muon.  ``Unidentified'' means that only one isolated lepton has
been positively identified in the event.
(For details see the description of cut~3 in 
Appendix I.3 of~\protect\cite{ppe124}.)
}} }
\label{tab-dlept}
\end{table}
For the  same event sample,
figure~\ref{fig-xlept} shows the distributions 
of~(a)  the energy scaled by the beam energy
 and~(b) the value of $-q \cos\theta$ of each
 charged lepton candidate.
The data, shown as points with error bars, are compared with the \smc\ predictions,
which are dominated by the final state \llnunu .

\begin{figure}[htbp]
 \epsfxsize=15.5cm
 \epsffile{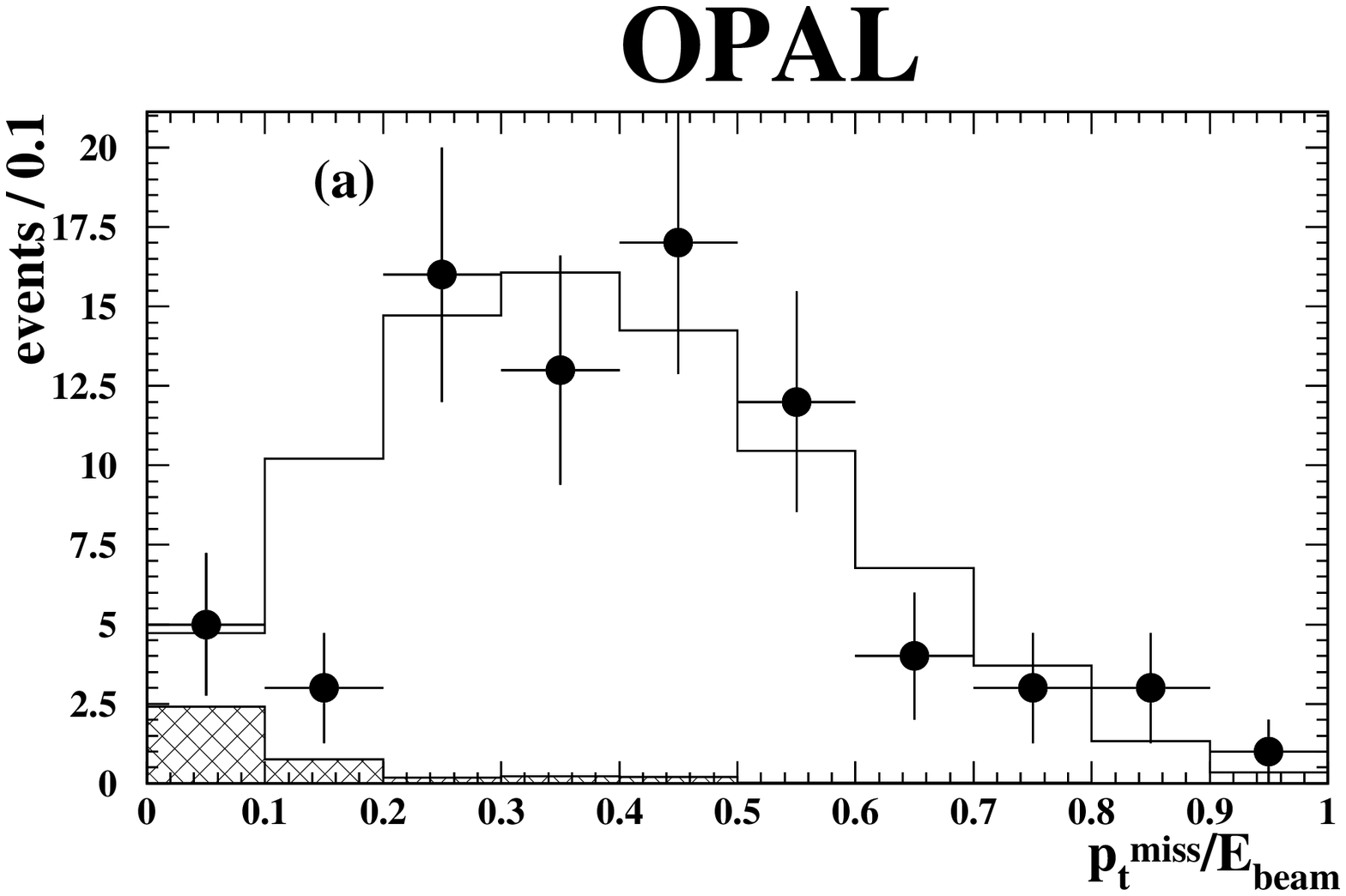}
 \epsfxsize=15.5cm
 \epsffile{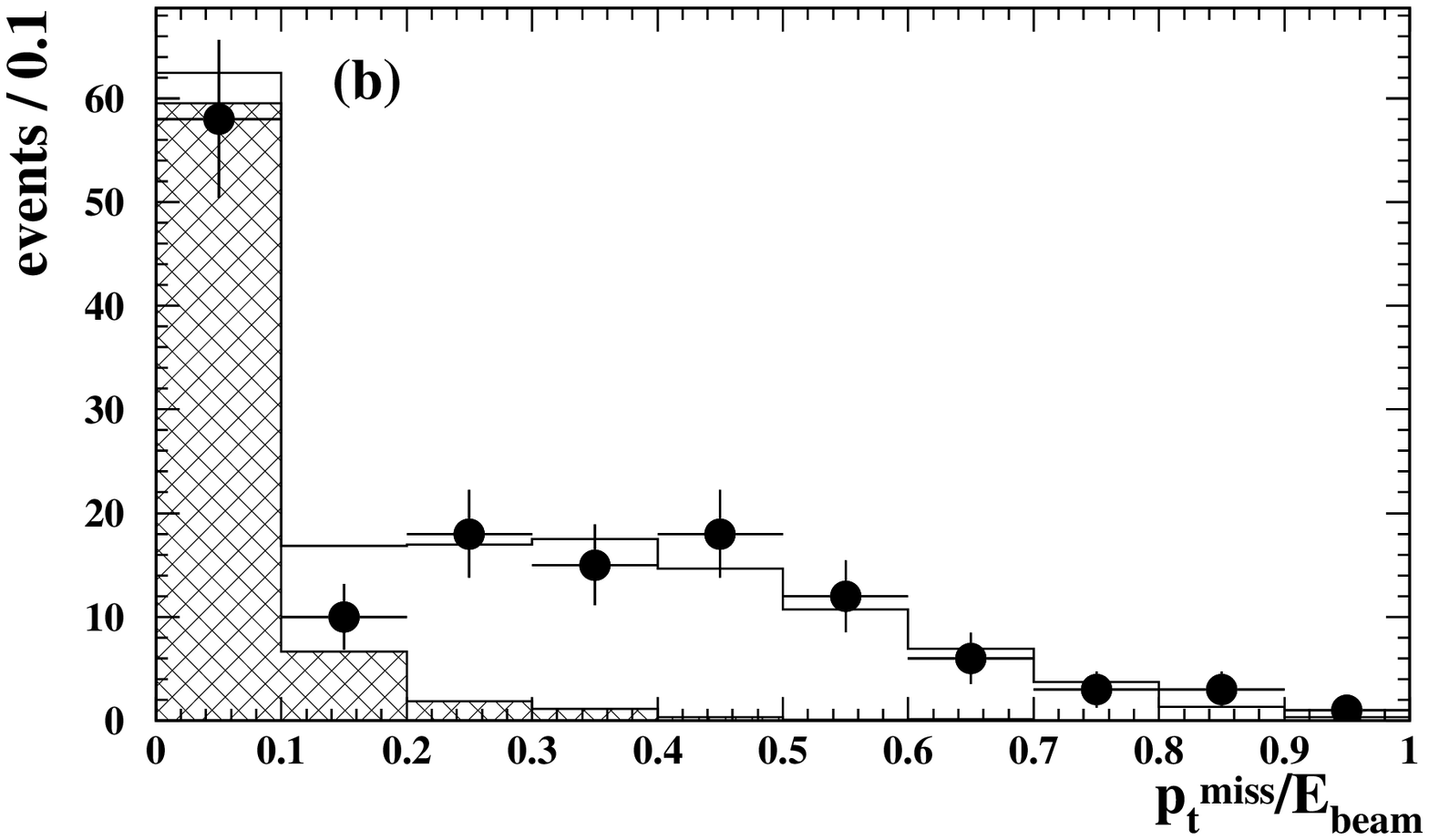}
 \caption{\sl 
(a) Distribution  of \stevt\  for
the general selection at \seight . 
(b) Distribution  of \stevt\  at \seight\ for the  event sample produced
 by relaxing some of the selection cuts
as described in appendix~I.3 of~\protect\cite{ppe124}.
The data are shown as the points with error bars.
The \smc\ prediction for \llnunu\  is shown as the
open histogram and the
background, arising mainly from processes with four charged leptons in
the final state, is shown as the cross-hatched histogram.
} 
\label{fig-stevt}
\end{figure}

Figure~\ref{fig-stevt}~(a) shows the distribution of \stevt\ 
for the general selection at \seight .
As a test of the degree to which the \smc\ describes the  \eell\ background,
figure~\ref{fig-stevt}~(b) shows the distribution of \stevt\ after
we relax some of the event selection cuts,
as described in appendix~I.3 of~\cite{ppe124}.
With these relaxed cuts
the number of observed candidates at \seight\ is 145; 
this  is consistent with the 
$149.3\pm2.0$ events expected from \sm\ sources, of which
$66.3\pm1.9$ arise from the final state \eell .
In each of the above checks the data
 are consistent with the Standard Model expectations.

The cuts used to veto two-photon background  introduce an inefficiency
in the event selection due to random detector occupancy
(principally in the SW, FD and MIP plug detectors)
that is not modelled in the \mc.
This inefficiency has been measured using randomly
triggered events collected during normal data taking.
In the data collected at \seight\
the inefficiency decreases from a value of 8.2\% for events
with very low missing transverse momentum to a negligible value for events
with $\stevt > 0.25$.
At $\sqrt{s} = 161$ and~172~GeV the backgrounds from
 off-momentum electrons in LEP were
much lower and the veto inefficiency was around 3\% for events
with very low missing transverse momentum.
When quoting expected numbers of \sm\ events and selection 
efficiencies for potential new physics sources, the
variation of veto inefficiency with \ptevt\ is taken into account.

\section{Additional Selection Criteria for New Particle Searches}
\label{sec:addkin}

Starting from the general selection of 
events containing two charged leptons and 
missing transverse momentum that was
discussed in section~\ref{sec:gen}, we search for the production of
new particles by applying
additional cuts
to suppress \sm\ sources of such events, the most important of which
are \llnunu\ and \eell .

The \sml\ from \wpair\ are characterised by the production of two leptons,
both with \spp\ around 0.5.
Equal numbers of $\mathrm{e}^\pm$, $\mu^\pm$ and $\tau^\pm$ are produced and
there is no correlation between the flavours of the two charged
leptons in the event.
In the \sm\ \eell\ events the two observed leptons tend to 
have low momentum. 

In the signal events the momentum distribution of the expected 
leptons varies strongly
as a function of the mass difference, \dm\ between the parent particle (e.g.,
selectron) and the invisible daughter particle (e.g., lightest neutralino),
and, to a lesser extent, $m$, the mass of the parent particle.
Slepton and charged Higgs boson pairs decay to produce two charged leptons
of the same flavour.
When performing a search for a particular new particle at a
particular point in $m$ and \dm , an event is considered as a potential
candidate only if the properties of the observed leptons are
consistent with expectations for 
signal events at those values of $m$ and \dm .

The method employed here differs from that described in~\cite{ppe124}
in two respects:
\begin{enumerate}
\item 
Use of the $-q \cos\theta$ of the two lepton candidates to discriminate
against the \smb\ in addition to the information on the 
lepton identification and energy of
the lepton candidates. 
\item 
Use of a likelihood technique to combine information from the
various discriminating variables.
\end{enumerate}

\begin{figure}[htbp]
 \epsfxsize=15.5cm
 \epsffile{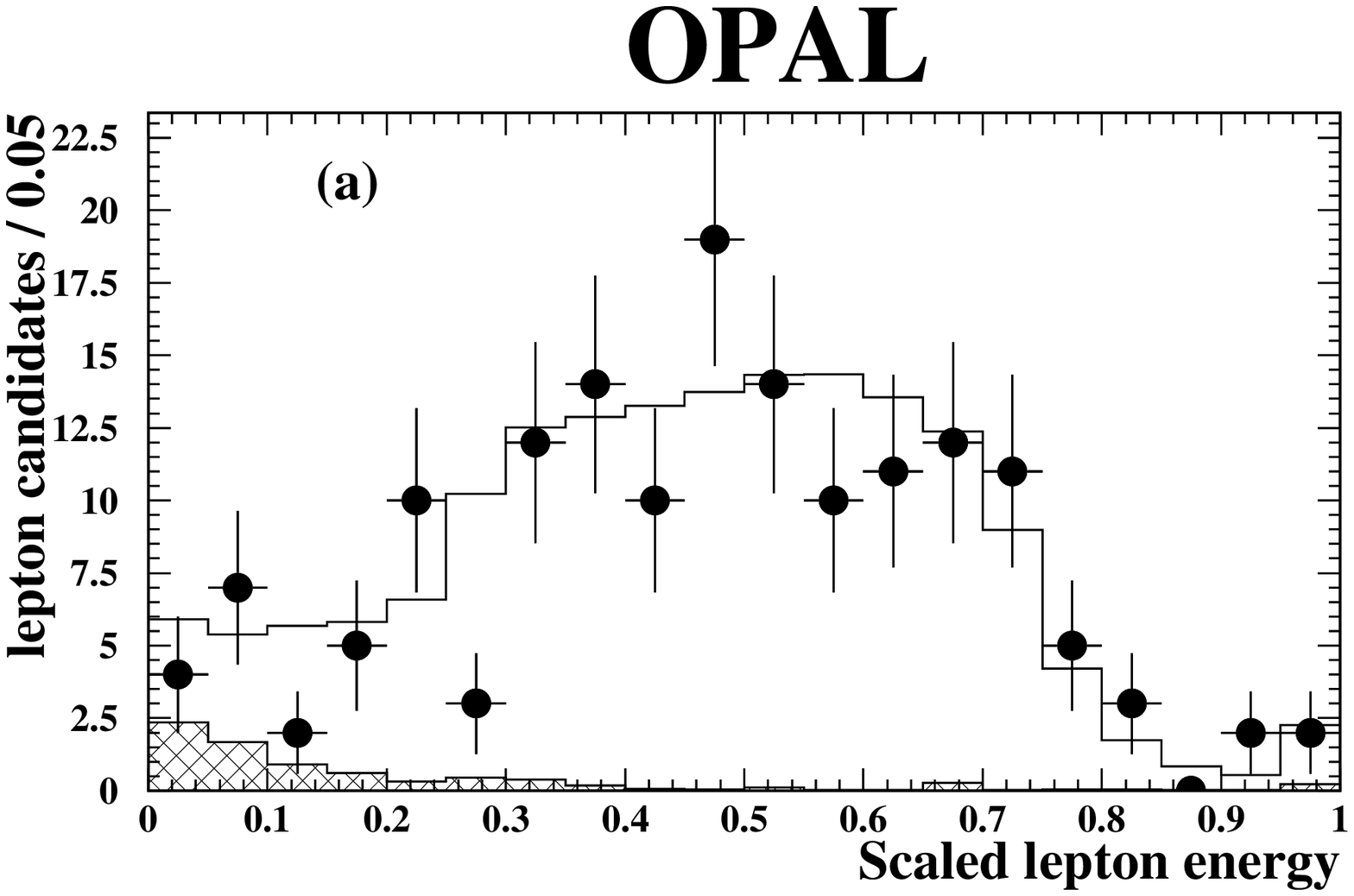}
 \epsfxsize=15.5cm
 \epsffile{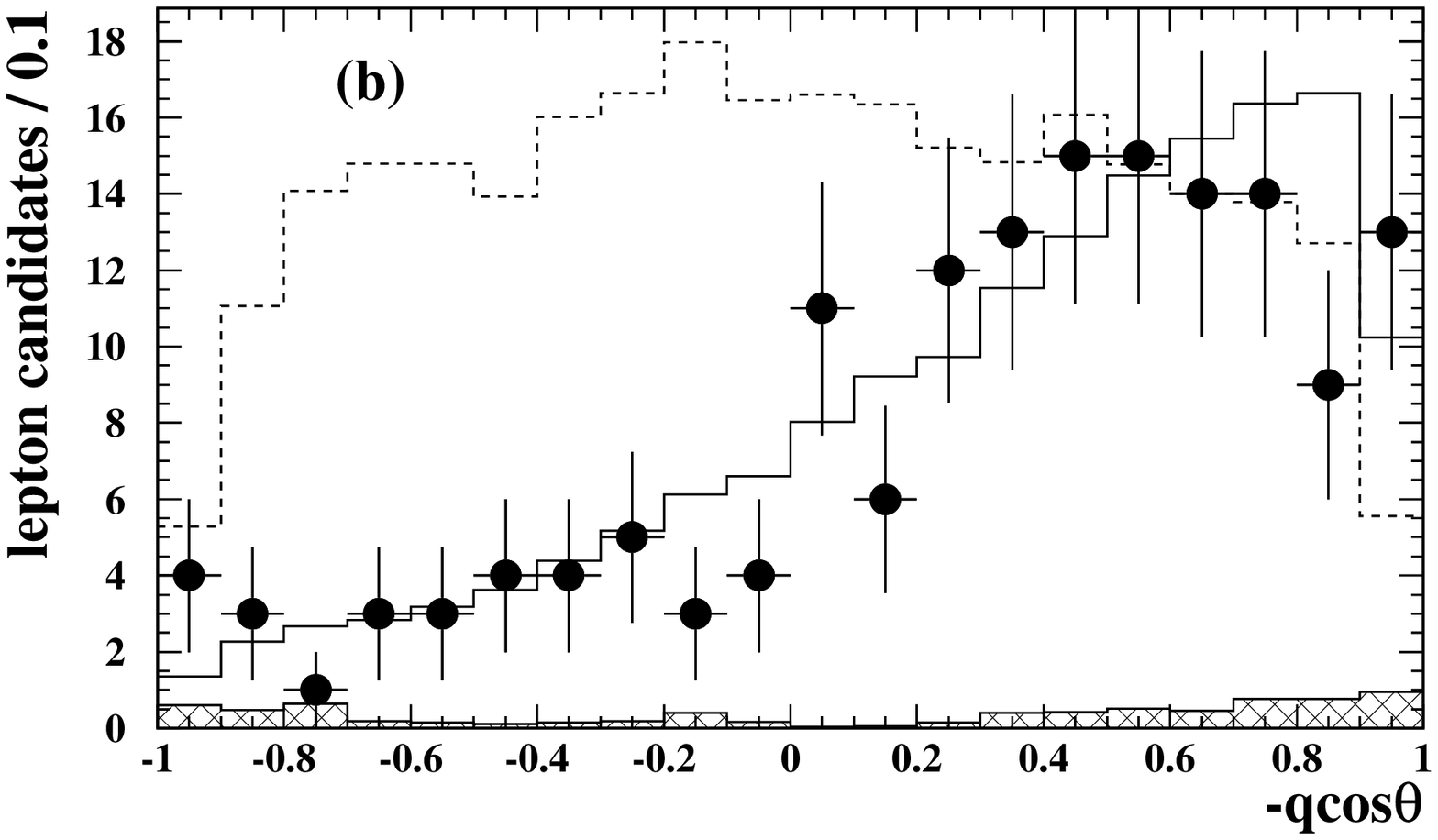}
 \caption{\sl Distributions of (a) the lepton energy divided by the beam
   energy, and~(b) $-q \cos\theta$ for
the general selection at \seight .  
The data are shown as the points with error bars.
The \smc\ prediction for \llnunu\  is shown as the
open histogram and the
background, arising mainly from processes with four charged leptons in
the final state, is shown as the cross-hatched histogram.
In~(b) the dashed histogram corresponds to the distribution expected
from smuon pair production.
\label{fig-xlept}
} 
\end{figure}

As is demonstrated by figure~\ref{fig-xlept}~(b), 
the distribution of $-q \cos\theta$ of the lepton
candidates in \wpair\ events is strongly forward peaked.
This may be compared with the dashed histogram in
figure~\ref{fig-xlept}~(b), which corresponds to the distribution expected
from smuon pair production and decay.
The distribution for smuons is symmetric
about $\cos{\theta}=0$.
This is because the smuon is a scalar; the angular distribution for
$s$-channel production of smuon pairs is expected to be
  proportional to $\sin^2\theta$ and 
the decay muon is expected to be isotropic 
     in the smuon rest frame.
In the search for pair production of smuons, staus and charged Higgs 
bosons the value of
$-q \cos\theta$ of the two lepton candidates is used to discriminate
against the \smb\ in addition to the information on the 
lepton identification and energy of
the lepton candidates. 

Selectrons may be produced via $t$-channel
neutralino exchange in addition to $s$-channel production.  This results
in the expected $- q \cos{\theta}$ distribution of selectrons being
model-dependent.
 Choices  of model 
parameters are possible for which the $t$-channel exchange is
dominant. 
In this case the
distribution will be forward-peaked similar to that of the \wpair\ background.
Similarly, charginos may be produced via $t$-channel sneutrino exchange.
 Therefore, the variable  $- q \cos{\theta}$
is not used in the discrimination of signal
and \sm\ background in the searches for selectrons and charginos.

\begin{figure}[htbp]
 \epsfxsize=\textwidth 
 \epsffile[0 0 580 600]{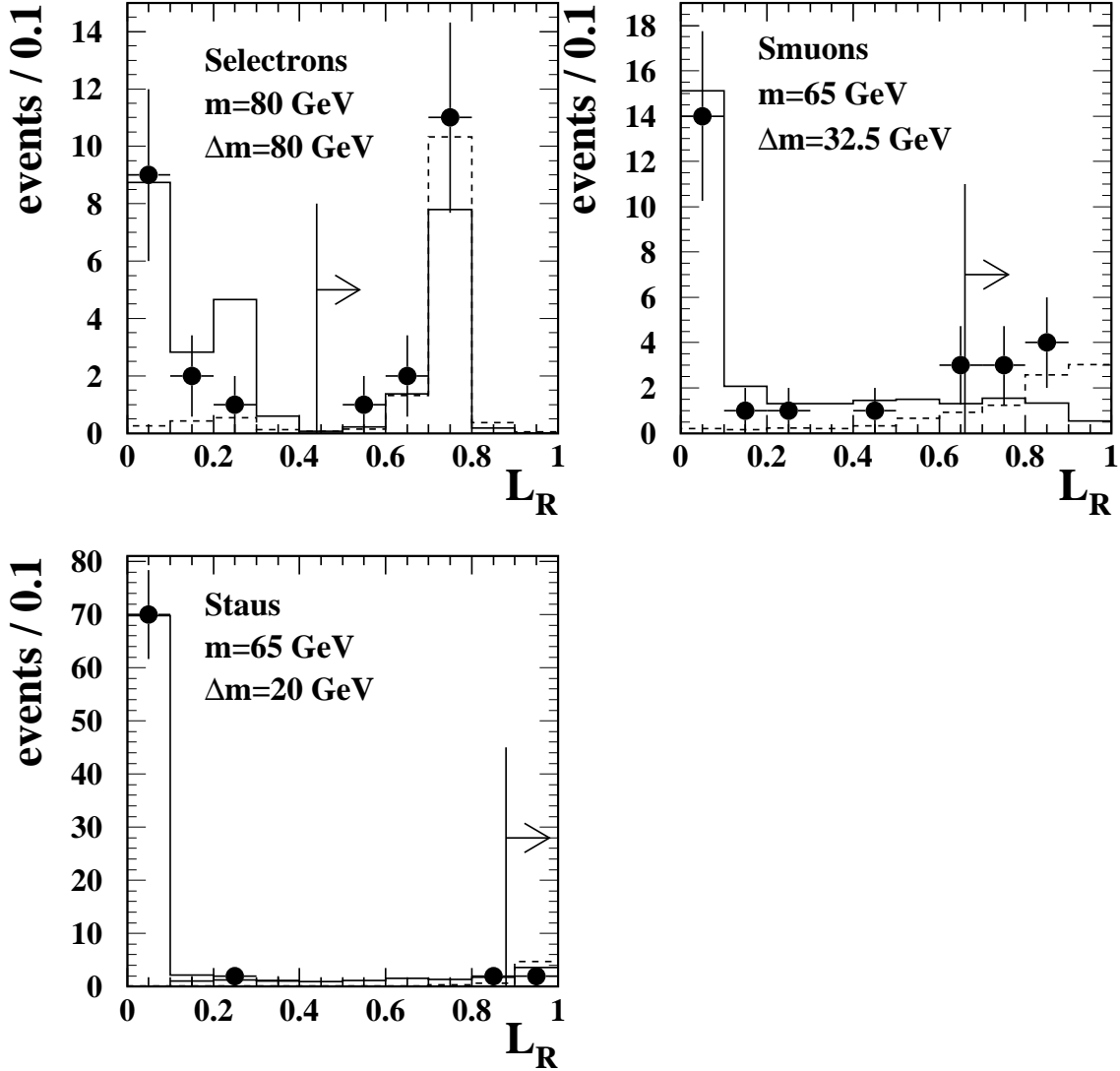}
 \caption{\sl 
Distributions of the quantity $\LR$, defined in 
section~\protect\ref{sec:addkin},
for selectrons, smuons and staus for some specific values of $m$ and \dm\ 
at $\protect\sqrt{s} = 183$~GeV.  The data are shown as the points with 
error bars.  The predicted \sm\ background is shown as the
solid histogram.  The dashed histogram corresponds to the distribution 
expected from selectron, smuon or stau pair production.  The signal
distribution is normalised in each case to represent the cross-section 
times branching ratio squared excluded at 
95\% CL by this analysis, for the values of $m$ and \dm\ shown.
} 
\label{fig:figLR}
\end{figure}

Discrimination  between 
\sm\ and new physics sources of lepton pair
events with missing momentum is performed by 
considering the likelihood that an event is consistent with being either signal
or background.
Given an event, for which the values of a set of variables $x_i$ are known, 
the likelihood\footnote{
Because the variables $x_i$ are correlated to some degree, $\LS$ is only an
approximation to the true likelihood.
However, since the signal efficiencies and expected \sm\ backgrounds are
estimated by applying the derived cuts to \mc\ samples no error is
introduced into the quoted results.
}, $\LS$, of the event being consistent with the signal hypothesis
is calculated as the product of the probabilities $\PS$ that the 
signal hypothesis
would produce an event with variable $i$ having value $x_i$:
$$\LS = \prod_{i}\PS.$$
The likelihood, $\LB$, of an event being consistent with the background
hypothesis is calculated similarly. 
The quantity $\LR$ is defined by:
$$\LR = \frac{\LS}{\LS+\LB}.$$
Event selection is
performed by making a cut on the value of $\LR$ rather than on the
individual variables $x_i$, which was the method used in~\cite{ppe124}.  
Distributions of $\LR$ for data and \sm\ 
\mc\ are compared for example values of $m$ and \dm\ at \seight , for 
selectrons, smuons and staus in figure~\ref{fig:figLR}.  

The optimisation of the selection cuts proceeds in a way similar to
that described in~\cite{ppe124}:
 the {\it a priori}\,
average value
of the 95\% CL upper limit on the cross-section for new physics 
is minimised by means of an automated procedure that makes use of Monte Carlo
samples of signal and Standard Model backgrounds, but not the
experimental data.
The optimisation is performed separately at each centre-of-mass energy.

\section{Numbers of Candidates, Backgrounds and Efficiencies}
\label{efficiency}

At example points in $m$ and \dm ,
table~\ref{tab-eff1} gives the number of selected events,
the number of events expected from \smp\ and the 
selection efficiency for new physics,   of the searches at \seight\
for  \selepair , \smupair\ and \staupair .
Table~\ref{tab-eff4} gives the same information for the searches for
\chargtwoo\ and \chargthreee .
The Monte Carlo statistical errors are given.

\begin{table}[htbp]
\centering
\begin{tabular}{||l||c|c|c||c|c|c||c|c|c||}
\hline\hline
       & \multicolumn{3}{c||}{selectrons}  
       & \multicolumn{3}{c||}{smuons}
       & \multicolumn{3}{c||}{staus} \\
\cline{2-10}
  \dm\ & \multicolumn{3}{c||}{\msele\ (GeV)}
       & \multicolumn{3}{c||}{\msmu\ (GeV)}
       & \multicolumn{3}{c||}{\mstau\ (GeV)} \\
\cline{2-10}
 (GeV) & 65 & 80 & 90 & 65 & 80 & 90 & 65 & 80 & 90 \\
\hline\hline
  \multicolumn{10}{||l||}{number of selected events} \\
\hline
2        &  0     &  0     &  0 &  0 &  0     &  0   &  0     &  0     &  0  \\
2.5      &  0     &  0     &  0 &  0 &  0     &  0   &  0     &  0     &  0  \\
5        &  1     &  0     &  0 &  0 &  0     &  0   &  1     &  1     &  1  \\
10       &  1     &  0     &  0 &  0 &  0     &  0   &  1     &  2     &  1  \\
20       &  0     &  0     &  0 &  2 &  0     &  0   &  3     &  1     &  1  \\
$m$/2    &  7     &  6     &  3 & 10 &  6     &  3   &  4     &  2     &  2  \\
$m$--20  & 15     & 11     &  4 &  9 &  9     &  6   &  4     &  4     &  3  \\
$m$--10  & 15     & 13     &  2 &  7 &  9     &  6   &  5     &  8     &  4  \\
$m$      & 15     & 14     &  2 &  4 &  9     &  6   &  5     &  9     &  4  \\
\hline\hline
\multicolumn{10}{||l||}{number of events expected from \smp} \\
\hline 
2        &
         0.5\p0.1 & 0.3\p0.1 & 0.1\p0.1 &
         0.1\p0.1 & 0.0\p0.0 & 0.0\p0.0 &
         0.1\p0.1 & 0.1\p0.1 & 0.1\p0.1 \\
2.5      &
         0.5\p0.1 & 0.3\p0.1 & 0.2\p0.1 &
         0.2\p0.1 & 0.1\p0.1 & 0.0\p0.0 &
         0.6\p0.2 & 0.3\p0.1 & 0.1\p0.1 \\
5        &
         0.6\p0.2 & 0.5\p0.2 & 0.1\p0.1 &
         0.3\p0.1 & 0.1\p0.1 & 0.1\p0.1 &
         1.8\p0.3 & 1.4\p0.2 & 1.1\p0.2 \\
10       &
         0.4\p0.1 & 0.1\p0.1 & 0.0\p0.0 &
         0.4\p0.1 & 0.2\p0.1 & 0.0\p0.0 &
         2.6\p0.3 & 2.2\p0.3 & 2.0\p0.3 \\
20       &
         1.7\p0.1 & 0.4\p0.0 & 0.1\p0.0 &
         1.8\p0.1 & 0.6\p0.1 & 0.1\p0.0 &
         3.9\p0.3 & 2.1\p0.2 & 1.8\p0.2 \\
$m$/2    &
         5.3\p0.2 & 3.5\p0.1 & 0.9\p0.1 &
         3.9\p0.2 & 3.7\p0.2 & 1.3\p0.1 &
         5.8\p0.3 & 6.0\p0.3 & 5.1\p0.3 \\
$m$--20  &
        10.8\p0.2 & 7.7\p0.2 & 1.5\p0.1 &
         4.6\p0.2 & 5.2\p0.2 & 1.7\p0.1 &
         7.3\p0.3 & 9.1\p0.4 & 6.7\p0.3 \\
$m$--10  &
        10.9\p0.2 & 9.2\p0.2 & 1.6\p0.1 &
         4.0\p0.2 & 5.2\p0.2 & 1.8\p0.1 &
         7.7\p0.3 & 9.6\p0.4 & 6.1\p0.3 \\
$m$      &
        11.1\p0.2 & 9.5\p0.2 & 1.8\p0.1 &
         3.0\p0.1 & 5.6\p0.2 & 2.2\p0.1 &
         7.6\p0.3 &10.1\p0.4 & 6.6\p0.3 \\
\hline\hline
  \multicolumn{10}{||l||}{selection efficiency (\%)} \\
\hline
2       &  8\p 1 &  3\p 1 &  1\p 0 & 10\p 1 &  2\p 0 &  0\p 0 &
  0\p 0 &  0\p 0 &  0\p 0 \\
2.5     & 24\p 1 & 18\p 1 & 12\p 1 & 24\p 1 & 15\p 1 & 11\p 1 &
  0\p 0 &  0\p 0 &  0\p 0 \\
5       & 51\p 2 & 53\p 2 & 54\p 2 & 56\p 2 & 55\p 2 & 57\p 2 &
 10\p 0 &  8\p 0 &  8\p 0 \\
10      & 64\p 1 & 60\p 1 & 64\p 1 & 67\p 1 & 70\p 1 & 71\p 1 &
 25\p 1 & 25\p 1 & 24\p 1 \\
20      & 65\p 1 & 69\p 1 & 69\p 1 & 70\p 1 & 76\p 1 & 76\p 1 &
 36\p 1 & 34\p 1 & 33\p 1 \\
$m$/2   & 63\p 1 & 66\p 1 & 68\p 1 & 63\p 1 & 70\p 1 & 75\p 1 &
 41\p 1 & 46\p 1 & 48\p 1 \\
$m$--20 & 70\p 1 & 69\p 1 & 67\p 1 & 63\p 1 & 68\p 1 & 68\p 1 &
 44\p 1 & 51\p 1 & 47\p 1 \\
$m$--10 & 72\p 1 & 74\p 1 & 64\p 1 & 61\p 1 & 67\p 1 & 66\p 1 &
 43\p 1 & 51\p 1 & 44\p 1 \\
$m$     & 73\p 1 & 74\p 1 & 67\p 1 & 55\p 1 & 68\p 1 & 71\p 1 &
 43\p 1 & 52\p 1 & 45\p 1 \\
\hline\hline
\end{tabular}
\caption[]{\sl
  \protect{\parbox[t]{15cm}{
Search for slepton pairs at \seight .
The number of selected events,
the number of events expected from \smp\ and the 
selection efficiency for different 
values of \mslept\ and \dm .
The selection efficiency for \staupair\ is calculated for the case
that the decay \dstau\ produces unpolarized $\tau^\pm$.
}} }   
\label{tab-eff1}
\end{table}

\begin{table}[htbp]
\centering
\begin{tabular}{||l||c|c|c|||l||c|c|c||}
\hline\hline
         \multicolumn{4}{||c|||}{\chargtwo}  
       & \multicolumn{4}{c||}{\chargthree}  \\ 
\hline
  \dm\ & \multicolumn{3}{c|||}{\mch\ (GeV)} &
  \dm\ & \multicolumn{3}{c||}{\mch\ (GeV)} \\
\cline{2-4}\cline{6-8}
(GeV)& 65 & 80 & 90 &(GeV)& 65 & 80 & 90 \\
\hline\hline
  \multicolumn{8}{||l||}{number of selected events} \\
\hline
1.5       &  1     &  0     &  0  & 3        &  0     &  0     &  0    \\
2.5       &  0     &  0     &  0  & 5        &  0     &  0     &  0    \\
5         &  2     &  0     &  0  & 10       &  2     &  2     &  1    \\
10        &  3     &  1     &  1  & 20       &  3     &  3     &  2    \\
20        &  5     &  2     &  1  & $m$/2    &  3     &  3     &  1    \\
$(m-15)/2$& 13     & 16     &  4  & $m$--20  &  5     &  5     & 11    \\
$m-35$    & 37     & 43     & 29  & $m$--10  & 11     & 22     & 19    \\
          &        &        &     & $m$      & 51     & 54     & 27    \\  
\hline\hline
\multicolumn{8}{||l||}{number of events expected from \smp} \\
\hline 
1.5       &
          1.5\p0.2 & 0.1\p0.1 & 0.1\p0.1 &
3        &         0.6\p0.1 & 0.5\p0.1 & 0.6\p0.2 \\
2.5       &
          1.1\p0.2 & 0.8\p0.2 & 0.2\p0.1 &
5        &         1.5\p0.2 & 0.8\p0.2 & 0.7\p0.2 \\
5         &
          2.7\p0.3 & 1.2\p0.2 & 0.6\p0.2 &
10       &         4.1\p0.4 & 2.5\p0.3 & 1.6\p0.2 \\
10        &
          5.1\p0.4 & 1.6\p0.2 & 0.7\p0.1 &
20       &         7.0\p0.5 & 4.7\p0.4 & 2.3\p0.3 \\
20        &
         14.7\p0.5 & 4.3\p0.3 & 1.4\p0.2 &
$m$/2    &         9.9\p0.5 & 6.3\p0.4 & 4.5\p0.3 \\
$(m-15)/2$&
         22.1\p0.5 &20.5\p0.5 & 9.3\p0.3 &
$m$--20  &        13.4\p0.5 &14.9\p0.5 &17.7\p0.5 \\
$m-35$    &
         42.5\p0.6 &51.4\p0.7 &33.9\p0.6 &
$m$--10  &        21.5\p0.6 &27.7\p0.6 &25.7\p0.5 \\
          & & & &
$m$      &        53.5\p0.7 &60.2\p0.7 &30.4\p0.5 \\
\hline\hline
  \multicolumn{8}{||l||}{selection efficiency (\%)} \\
\hline
1.5       &  0\p 0 &  0\p 0 &  0\p 0 &
3        &  3\p 0 &  1\p 0 &  1\p 1 \\
2.5       & 14\p 1 &  7\p 0 &  5\p 1 &
5        & 15\p 1 & 13\p 1 & 10\p 1 \\
5         & 45\p 2 & 34\p 1 & 39\p 1 &
10       & 39\p 1 & 38\p 1 & 38\p 1 \\
10        & 54\p 2 & 44\p 2 & 45\p 2 &
20       & 48\p 2 & 49\p 2 & 45\p 2 \\
20        & 60\p 1 & 62\p 1 & 58\p 2 &
$m$/2    & 51\p 2 & 51\p 2 & 53\p 2 \\
$(m-15)/2$&  ---   & 67\p 1 & 67\p 1 &
$m$--20  & 51\p 2 & 54\p 2 & 65\p 1 \\
$m-35$    & 65\p 1 & 73\p 1 & 70\p 1 &
$m$--10  & 53\p 2 & 61\p 1 & 64\p 1 \\
          & & & &
$m$      & 67\p 1 & 74\p 1 & 71\p 1 \\
\hline\hline
\end{tabular}
\caption[]{\sl
  \protect{\parbox[t]{15cm}{
Search for chargino pairs 
(2-body decays: \dchtwo\  and 3-body decays: \dchthree ) at \seight .
The number of selected events,
the number of events expected from \smp\ and the 
selection efficiency for different 
values of \mch\ and \dm .
}} }   
\label{tab-eff4}
\end{table}

It should be noted that an individual candidate event may be
consistent with a given new physics hypothesis over a range of $m$ and
\dm\ values.
Similarly, an individual candidate event may be
consistent with more than one new physics hypothesis and may,
therefore, give entries in more than one of the above tables.
At high \dm , the expected number of \sm\ events is made up almost entirely
(typically greater than 98\%) of \wpair\ events, whereas at low \dm ,
it is typical for approximately half of the \sm\ events to
arise from \wpair\ events, with the remainder arising predominantly
from two-photon events.

For each search channel, table~\ref{tab-cand} shows the total number
of selected candidates over all values of $m$ and \dm\ compared with
the \sm\ expectations.
\begin{table}[htbp]
\centering
\begin{tabular}{||l||c|c|c|c|c|c||}
\hline\hline
  Search & \multicolumn{6}{c||}{$\roots$ (GeV)} \\
\cline{2-7}
Channel & \multicolumn{2}{c|}{161} & \multicolumn{2}{c|}{172} & \multicolumn{2}{c||}{183} \\
\cline{2-7}
        & data & SM & data & SM & data & SM \\
\hline\hline
\selepair                    & 1 & 1.0 & 2 & 2.2 & 16 & 14.0 \\
\smupair                     & 1 & 0.8 & 2 & 2.0 & 13 & 11.5 \\
\staupair                    & 2 & 1.4 & 4 & 2.9 & 11 & 13.0 \\
$\mathrm{H}^+\mathrm{H}^-$   & 1 & 1.0 & 4 & 2.6 & 9 & 11.5 \\
\chargthree                  & 3 & 3.3 & 7 & 8.4 & 62 & 71.0 \\
\chargtwo                    & 2 & 2.9 & 7 & 7.3 & 48 & 58.2 \\
\hline\hline
\end{tabular}
\caption[]{\sl
  \protect{\parbox[t]{15cm}{
For each search channel at each value of $\roots$, the total number
of selected candidates over all values of $m$ and \dm\ compared with
the \sm\ expectations.
}} }   
\label{tab-cand}
\end{table}

In general, the \smc\ provides a good description of the data in
tables~\ref{tab-eff1}--\ref{tab-cand}.
The most significant difference between data and \mc\ 
is seen in table~\ref{tab-eff1} for smuons at
large \dm .  The excess is most marked at $m$ = 65~GeV,
\dm\ = $m$/2.
It is difficult to assess quantitatively the significance
of this excess, given that
there are strong correlations in the selected event samples:
a) among individual bins for a particular search channel and b) among
the six different search channels.

Slepton pair efficiencies are evaluated for right-handed sleptons, both 
decaying to lepton and lightest 
neutralino $\sell^\pm \rightarrow  {\ell^\pm} \nt_1$. 
The slepton pair events were generated at $\mu = -200$~GeV and
$\tan{\beta} = 1.5$ using {\sc Susygen}.
The selection efficiency for selectrons depends on
the angular distribution of the produced selectrons
and this will depend on the size of the neutralino-mediated $t$-channel
contribution to the cross-section. 
We have found by varying $\mu$ and $\tan{\beta}$ that the above choice
gives a conservative estimate of the selection efficiency. 
In the generation of the stau pair events  using {\sc Susygen}
the produced taus are unpolarized.
 As noted in~\cite{taupol}
 the polarisation of taus in the decay of staus 
 can vary from $+1$ to $-1$ depending on the neutralino field content.
 This leads to a model dependence of the 
 expected momentum spectrum of the visible tau decay products. 

The experimental signature of 
\hpair\ production followed by the decay
$\mathrm{H}^{\pm} \rightarrow \tau^\pm \nu_\tau$ is similar to that
of \staupair\ production, 
$ \tilde{\tau}^+
\tilde{\tau}^-\to  \tau^+ \tilde{\chi}^0_1 \tau^- \tilde{\chi}^0_1$, 
for the case
when the $\tilde{\chi}^0_1$ is massless and
stable.
The selection efficiencies for \hpair\ events generated using 
{\sc Hzha} and {\sc Pythia} are
consistent with those for \staupair\ events with
$m_{\tilde{\chi}^0_1} = 0$ generated using {\sc Susygen}\footnote{
Note that in {\sc Hzha} the tau polarisation in the decay \dH\ is
handled correctly, whereas in  {\sc Susygen} the taus are unpolarised.
The average selection efficiencies obtained with the two \mc s are
consistent within the 3\% statistical accuracy of the comparison.
This represents a check of the model dependence also of the \staupair\
selection efficiencies at high \dm .
}.
Because the available Monte Carlo samples for the latter process have
higher statistics, we use them to evaluate the selection efficiencies
for \hpair .  

Monte Carlo samples for \chpair\ production followed by the decay:
$\chpm \rightarrow \ell^\pm \snu$ (2-body decay) are generated 
using {\sc Susygen}.
Given the limit of 37.1~GeV on the mass of the lightest sneutrino 
from LEP1~\cite{snu},
\chargtwo\ events are not generated for sneutrino masses less than
35~GeV.
Monte Carlo samples for \chpair\ production in which the chargino 
decays via a virtual or real W, \dchthree,  (3-body decays)
are generated using   {\sc
  Dfgt} at \seight\ and {\sc Susygen} at $\sqrt{s} = 161$ and~172~GeV.

It can be seen that sizable selection efficiencies have been obtained
for \selepair , \smupair\ and  \chargtwo , even when \dm\ is as low as
2~GeV.
However, for \staupair\ and  \chpair\ (3-body decays) there are additional
invisible particles (neutrinos) in the final state.
 The visible
leptons are therefore less energetic and the selection efficiencies at
low \dm\ values are reduced.

\section{New Particle Search Results}
\label{results}

The number of observed candidate events 
and their kinematic properties are
compatible with the 
expectations from Standard Model processes. 
We present limits on the pair production
of charged scalar leptons, leptonically decaying charged Higgs bosons 
and charginos that decay to produce a charged lepton and invisible particles.
The sensitivity of the searches is significantly improved with respect
to~\cite{ppe124} because of: the increased centre-of-mass energy and
integrated luminosity; the reduced \eell\ background achieved due to 
the detector upgrade (MIP plug); and the
improved analysis techniques (use of  $-q \cos\theta$ and likelihood
techniques). 

As described in section~\ref{sec:addkin}, 
the additional event selection cuts 
for a given search channel vary as a function of  $m$ and \dm .
As input to the limit calculation for each new particle
search  we calculate at each value of $m$, \dm , and $\sqrt{s}$:
the number of observed candidates, the number of expected events from
\smb , the selection efficiency, and the pair production cross-section
relative to that at \seight . 
The 95\% CL upper limits at \seight\ are
obtained by combining the data at the three e$^+$e$^-$ centre-of-mass 
energies 161, 172 and 183~GeV using
the Likelihood Ratio method~\cite{LR}.

The number of selected candidate events is determined at each
kinematically allowed point on a 0.2~GeV by 0.2~GeV grid of $m$ and \dm .
Monte Carlo signal events are available only at certain particular
 values of $m$ and \dm .
The values of $m$ range typically from $m = 45$~GeV up to $m \approx \Ebeam$ in
5~GeV steps.
The values of \dm\ correspond to those
given in tables~\ref{tab-eff1} and~\ref{tab-eff4}.
Signal efficiencies at intermediate values of $m$ and \dm\ are
obtained by linear 2-dimensional interpolation.
In addition to the \mc\ statistical error, we assign a 5\% 
systematic error on the estimated selection efficiency to take into
account uncertainties in: trigger efficiency, detector
occupancy, lepton identification efficiency, luminosity measurement, 
interpolation procedure, and 
deficiencies in the \mc\ generators and the detector simulation.

At high values of \dm\ the dominant background to 
the searches for new physics results from \wpair\ production.
High statistics \mc\ samples for this process are available that describe well
the OPAL data~\cite{Wmass,wwpaper172,wwpaper}.
In addition to the \mc\ statistical error, we assign a 5\% 
systematic error on the estimated background to take into account 
deficiencies in the \mc\ detector simulation.
At low values of \dm\ the dominant background results from \eell\
events.
The background uncertainty at low \dm\ is dominated by the limited \mc\
statistics; the
uncertainty is typically 20--80\% at low \dm . 
In setting limits the \mc\ statistical errors and other
systematics are taken into account according to the 
method described in~\cite{cousins}.

\begin{figure}[htbp]
 \epsfxsize=\textwidth 
 \epsffile[0 0 580 600]{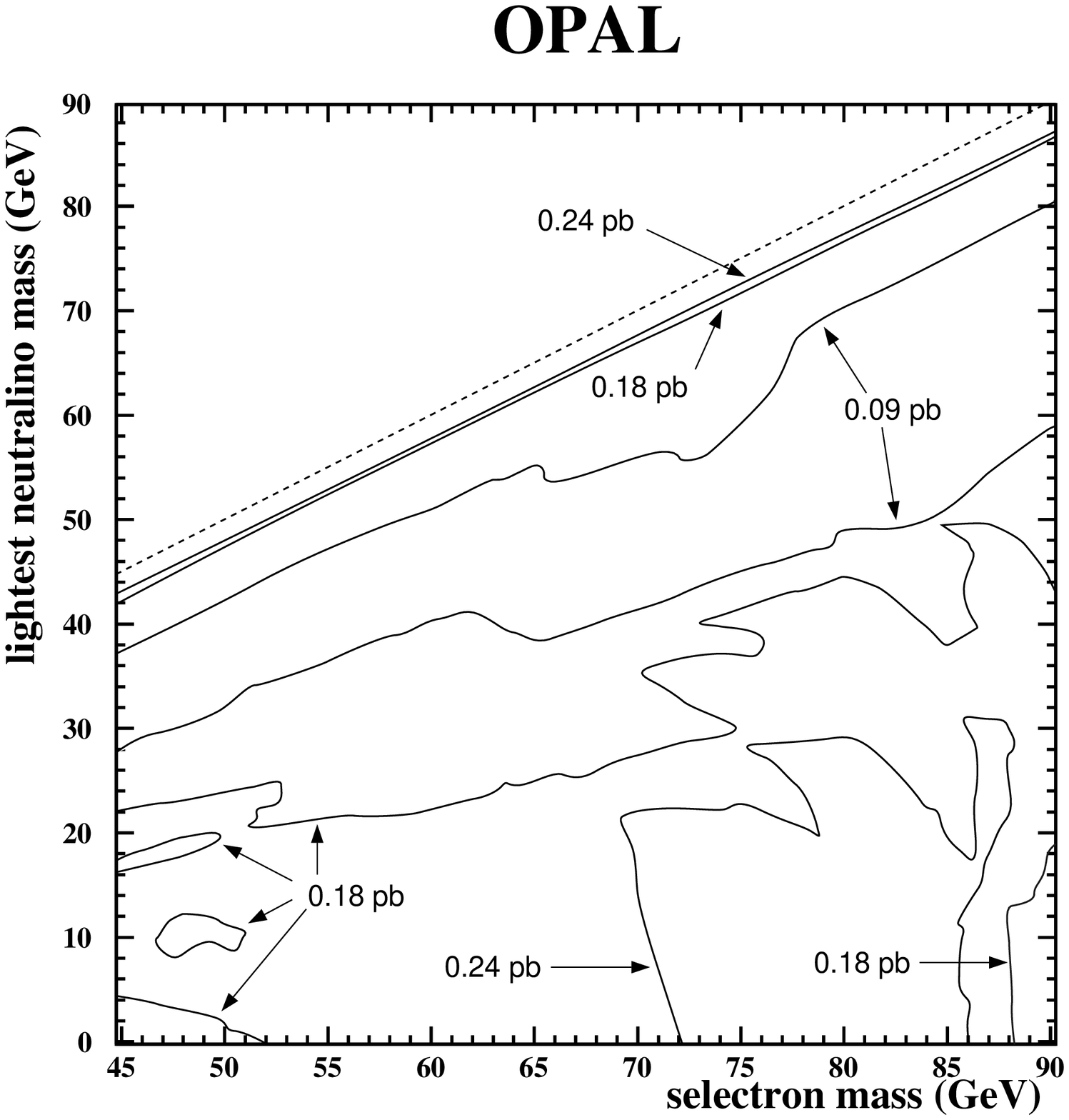}
 \caption{\sl 
Contours of the 95\% CL upper limits on the selectron pair
cross-section times $BR^2(\sele \rightarrow \mathrm{e} \nt_1)$
at $\protect\sqrt{s} = 183$~GeV
based on combining the $\protect\sqrt{s} = 161-183$~GeV data-sets 
assuming a $\beta^3/s$ dependence of the cross-section.
The kinematically allowed region is indicated by the dashed line.
} 
\label{fig:limit_1}
\end{figure}

\begin{figure}[htbp]
 \epsfxsize=\textwidth 
 \epsffile[0 0 580 600]{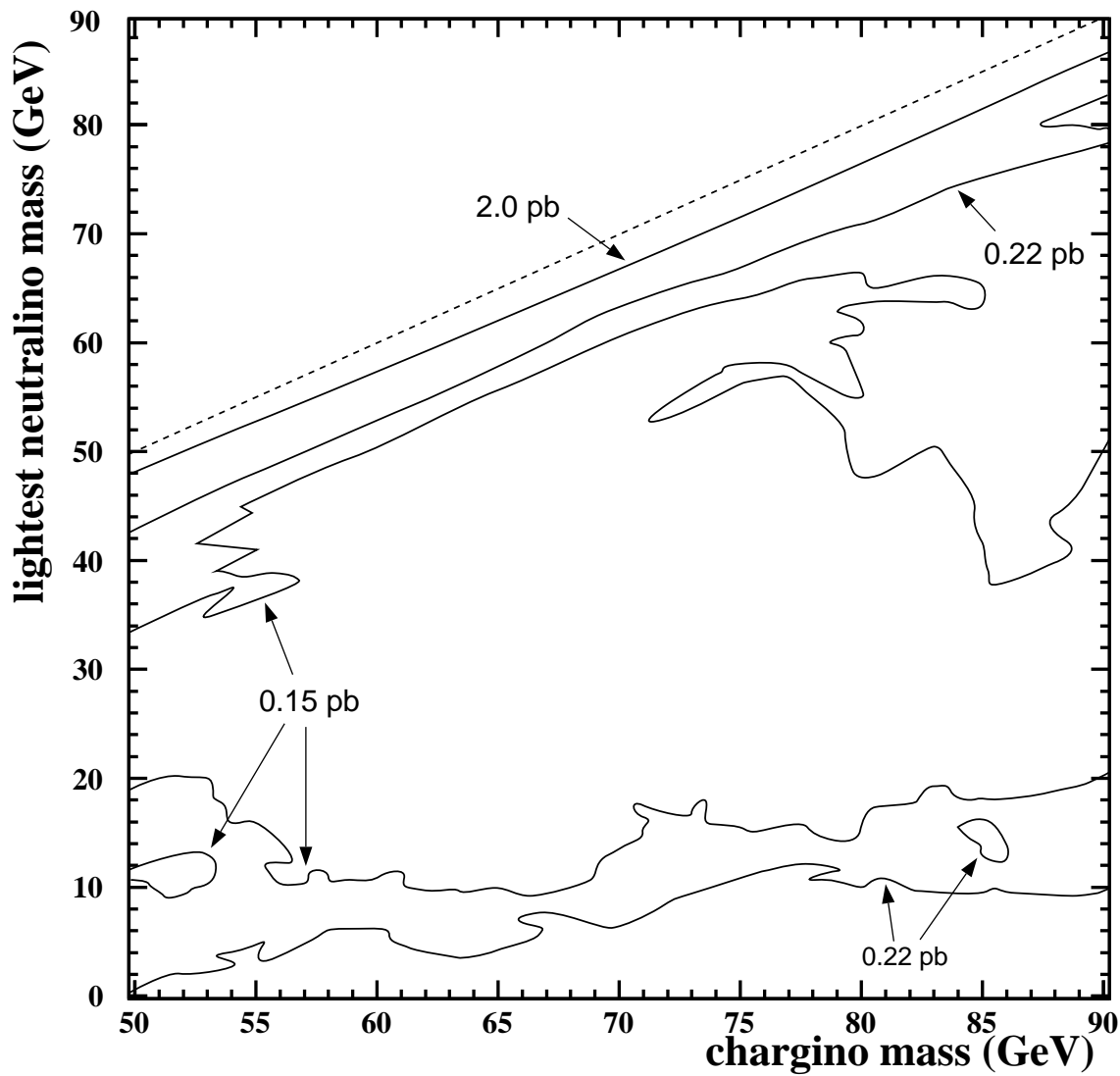}
 \caption{\sl 
Contours of the 95\% CL upper limits on the chargino pair
cross-section times branching ratio squared for 
$\chpm \rightarrow \ell^\pm \nu \chz$
 (3-body decay) at $\protect\sqrt{s}$~=~183~GeV,
The limit is obtained by combining the 161--183~GeV data-sets 
assuming a $\beta/s$ dependence of the cross-section.
The kinematically allowed region is indicated by the dashed line.
} 
\label{fig:limit_4}
\end{figure}

Limits on the production cross-section times branching ratio squared for new
physics processes are now presented in a manner intended to 
minimise the number of model assumptions.
The 95\% CL upper limits at \seight\ shown in 
figures~\protect\ref{fig:limit_1}~--~\protect\ref{fig:limit_4} are
obtained by combining the data at the three e$^+$e$^-$ centre-of-mass 
energies 161, 172 and 183~GeV using
the assumption that the
cross-section varies as $\beta^3/s$ for sleptons
and $\beta/s$ for charginos.
The chosen functional forms are used for simplicity in presenting the
data and represent an approximation, particularly
for processes in which $t$-channel exchange may be important, that is,
selectron pair and chargino pair production.
In these cases the cross-section
dependence on e$^+$e$^-$ centre-of-mass energy 
is model dependent, depending on the mass of the exchanged particles 
and the couplings of the neutralinos and charginos.

\begin{figure}[htbp]
 \epsfxsize=\textwidth 
 \epsffile[0 0 580 600]{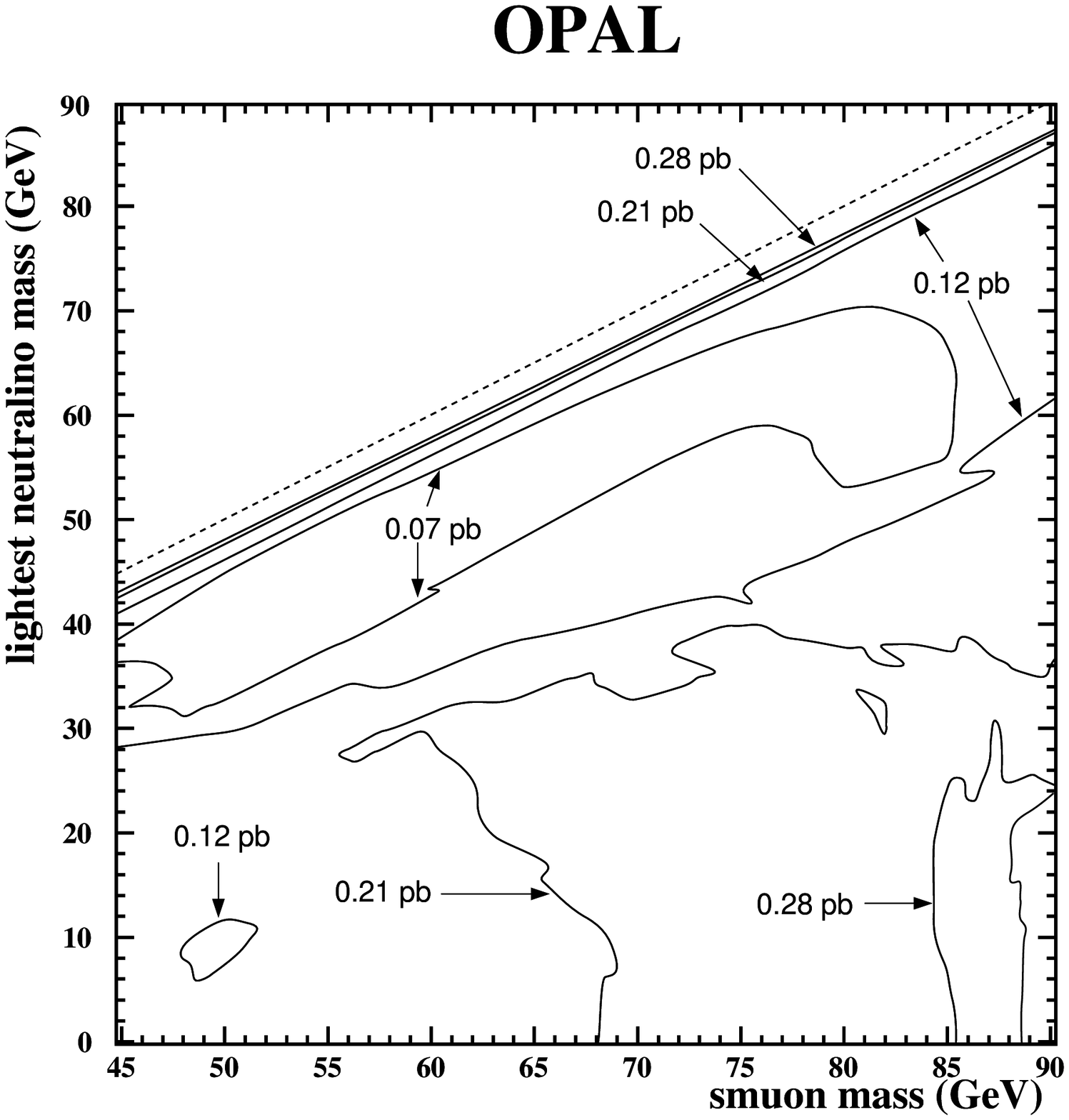}
 \caption{\sl 
Contours of the 95\% CL upper limits on the smuon pair
cross-section times $BR^2(\smu \rightarrow \mu \nt_1)$
at $\protect\sqrt{s} = 183$~GeV
based on combining the $\protect\sqrt{s} = 161-183$~GeV data-sets 
assuming a $\beta^3/s$ dependence of the cross-section.
The kinematically allowed region is indicated by the dashed line.
} 
\label{fig:limit_2}
\end{figure}

\begin{figure}[htbp]
 \epsfxsize=\textwidth 
 \epsffile[0 0 580 600]{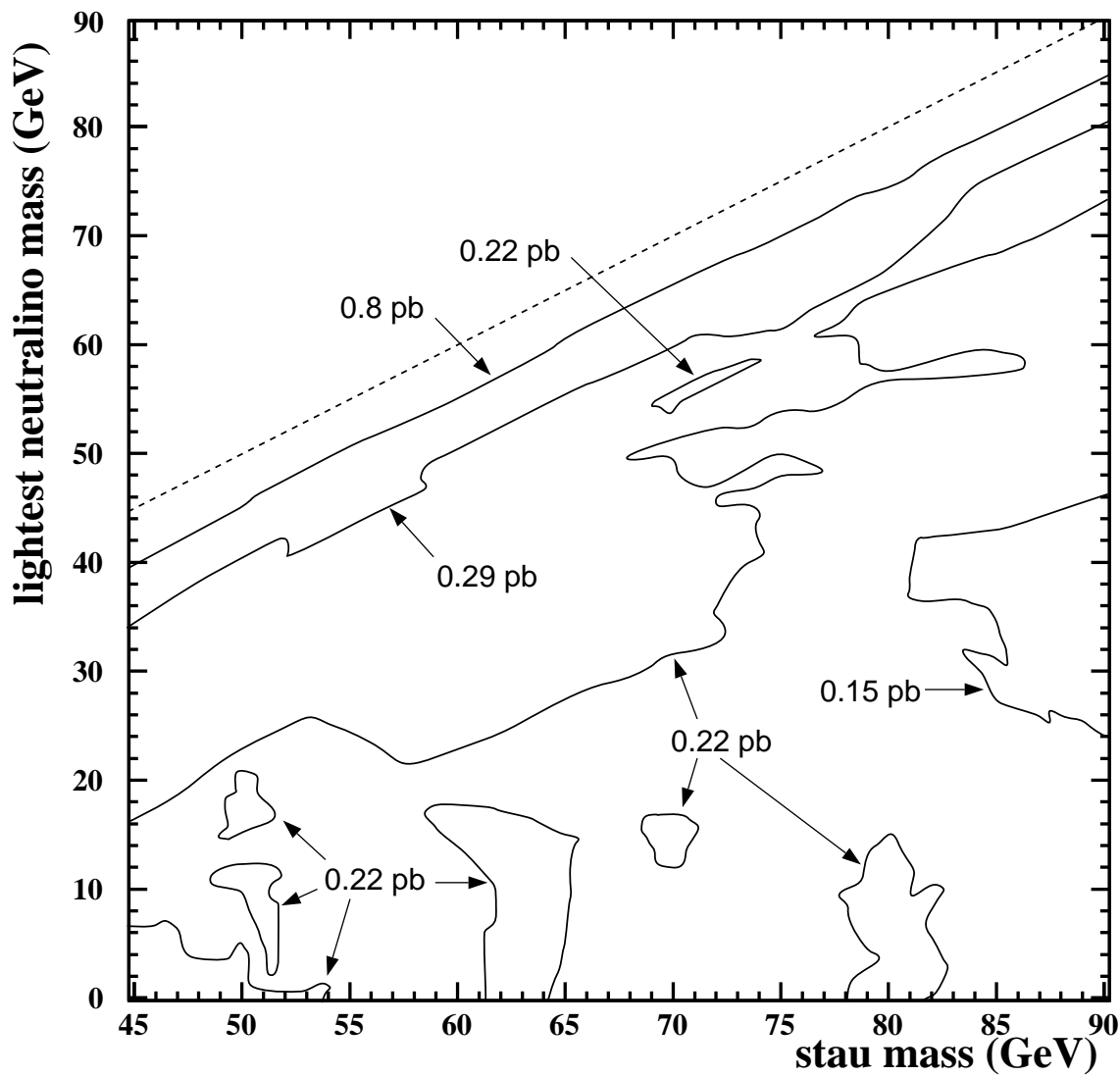}
 \caption{\sl 
Contours of the 95\% CL upper limits on the stau pair
cross-section times $BR^2(\stau \rightarrow \tau \nt_1)$
at $\protect\sqrt{s} = 183$~GeV
based on combining the $\protect\sqrt{s} = 161-183$~GeV data-sets 
assuming a $\beta^3/s$ dependence of the cross-section.
The selection efficiency for \staupair\ is calculated for the case
that the decay \dstau\ produces unpolarized $\tau^\pm$.
The kinematically allowed region is indicated by the dashed line.
} 
\label{fig:limit_3}
\end{figure}

Upper limits at  95\% CL on the selectron pair cross-section 
at~\seight\ times 
branching ratio squared for the decay \dsele\
are shown in figure~\ref{fig:limit_1} as a function of selectron mass 
and lightest neutralino mass.
These limits are applicable to
$\tilde{\mathrm e}^+_{\mathrm L}\tilde{\mathrm e}^-_{\mathrm L}$ and 
$\tilde{\mathrm e}^+_{\mathrm R}\tilde{\mathrm e}^-_{\mathrm R}$ production.
The corresponding plots for  the smuon and stau pair searches are shown in  
figures~\ref{fig:limit_2} and~\ref{fig:limit_3}, respectively.
Note that if the  LSP is the gravitino (effectively massless), then for 
 prompt slepton decays to lepton-gravitino the experimental signature 
would be the same as that
for  \dslept\ with a massless neutralino.
In this case the limits given in 
figures~\protect\ref{fig:limit_1}~--~\protect\ref{fig:limit_3}
for $m_{\tilde{\chi}^0_1} = 0$ may be interpreted as limits on the decay to 
lepton-gravitino.

\begin{figure}[htbp]
 \epsfxsize=\textwidth 
 \epsffile[0 0 580 600]{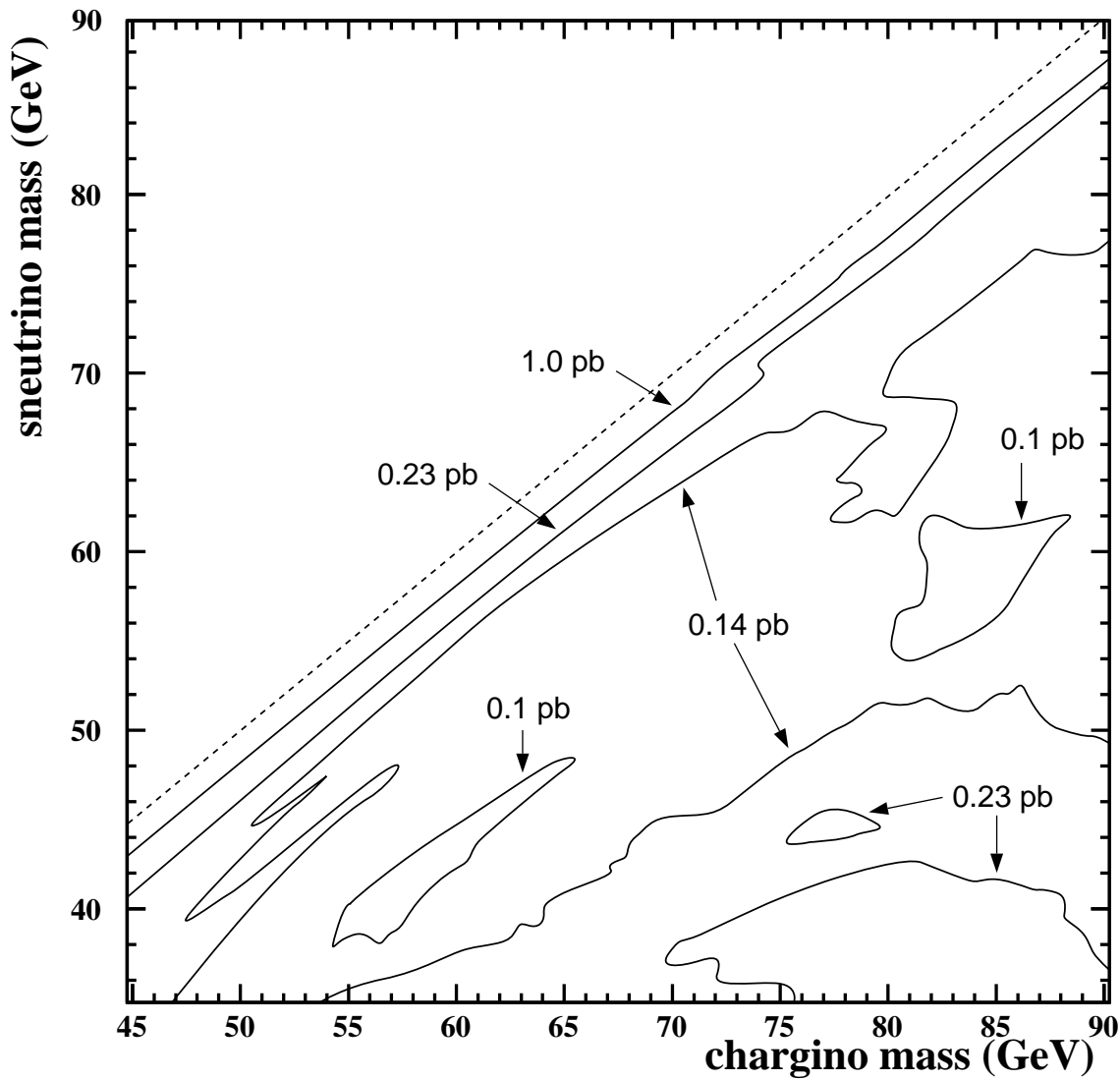}
 \caption{\sl 
Contours of the 95\% CL upper limits on the chargino pair
cross-section times branching ratio squared for 
$\chpm \rightarrow \ell^\pm \snu$ (2-body decay)
at $\protect\sqrt{s}$~=~183~GeV.
The  limits have been calculated for the 
case where the three sneutrino 
generations are mass degenerate.
The limit is obtained by combining the 161--183~GeV data-sets 
assuming a $\beta/s$ dependence of the cross-section.
The kinematically allowed region is indicated by the dashed line.
} 
\label{fig:limit_8}
\end{figure}

The upper limit at 95\% CL on the chargino pair production 
cross-section times branching
ratio squared for the decay \dchtwo\  (2-body decay)
is shown in figure \ref{fig:limit_8}. 
The  limit has been calculated for the 
case where the three sneutrino 
generations are mass degenerate.
The upper limit at 95\% CL on the chargino pair production 
cross-section times branching
ratio squared for the decay \dchthree\ (3-body decay)
is shown in figure \ref{fig:limit_4}. 
In a forthcoming paper~\cite{chargino} the search described here
for acoplanar di-lepton events will be combined with searches in
other final states to set more general limits on
chargino pair production.

\begin{figure}[htbp]
 \epsfxsize=\textwidth 
 \epsffile[0 0 580 600]{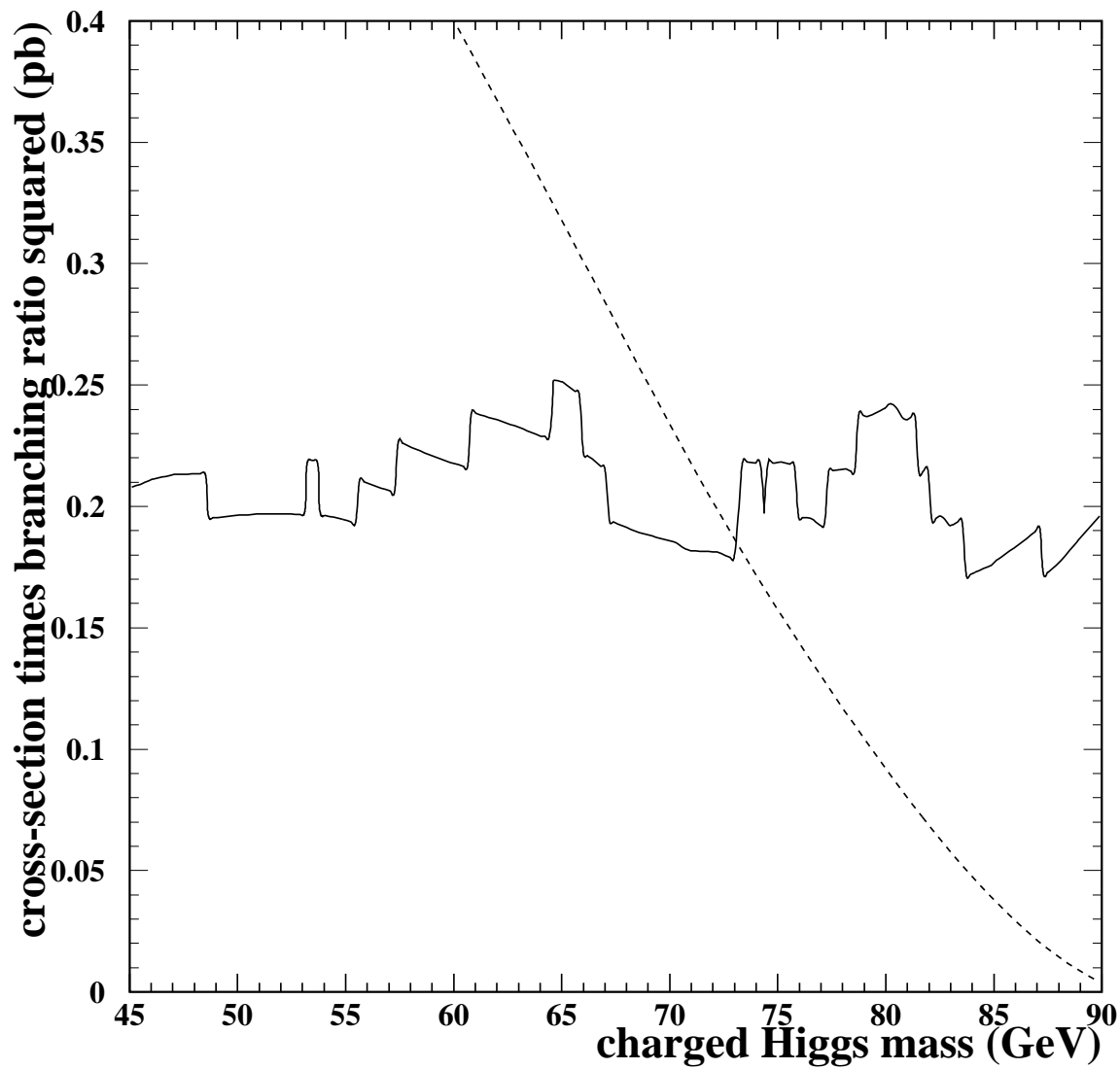}
 \caption{\sl 
The solid line shows the 95\% CL upper limit on the 
charged Higgs pair production 
cross-section times branching ratio squared for the decay \dH\
at $\protect\sqrt{s}$~=~183~GeV.
The limit is obtained by combining the 161--183~GeV data-sets 
assuming the \mH\ and $\protect\sqrt{s}$ 
dependence of the cross-section predicted
by {\sc Pythia}.
For comparison, the dashed curve shows the prediction from {\sc Pythia}
 at $\protect\sqrt{s}$~=~183~GeV
assuming a 100\% branching ratio for the decay \dH .
} 
\label{fig:limit_5}
\end{figure}

The upper limit at 95\% CL on the charged Higgs boson pair production 
cross-section times branching ratio squared for the decay \dH\
is shown as a function of \mH\ as the solid line in figure \ref{fig:limit_5}. 
The limit is obtained by combining the 161--183~GeV data-sets 
assuming the \mH\ and $\sqrt{s}$ 
dependence of the cross-section predicted
by {\sc Pythia}, which takes into account the effect of 
initial state radiation.
The dashed  line in figure \ref{fig:limit_5} 
shows the prediction from {\sc Pythia}
 at $\protect\sqrt{s}$~=~183~GeV
for a 100\% branching ratio for the decay \dH .
With this assumption we set a lower limit  at 95\% CL on \mH\ of 73~GeV.
In a forthcoming paper~\cite{higgs} the search described here
for acoplanar di-tau events will be combined with searches in
the final states $\tau\nbar\qpair$ and \qpair\qpair\ to set limits on
charged Higgs boson pair production for arbitrary \dH\ branching ratio. 

\begin{figure}[htbp]
 \epsfxsize=\textwidth 
 \epsffile[0 0 580 600]{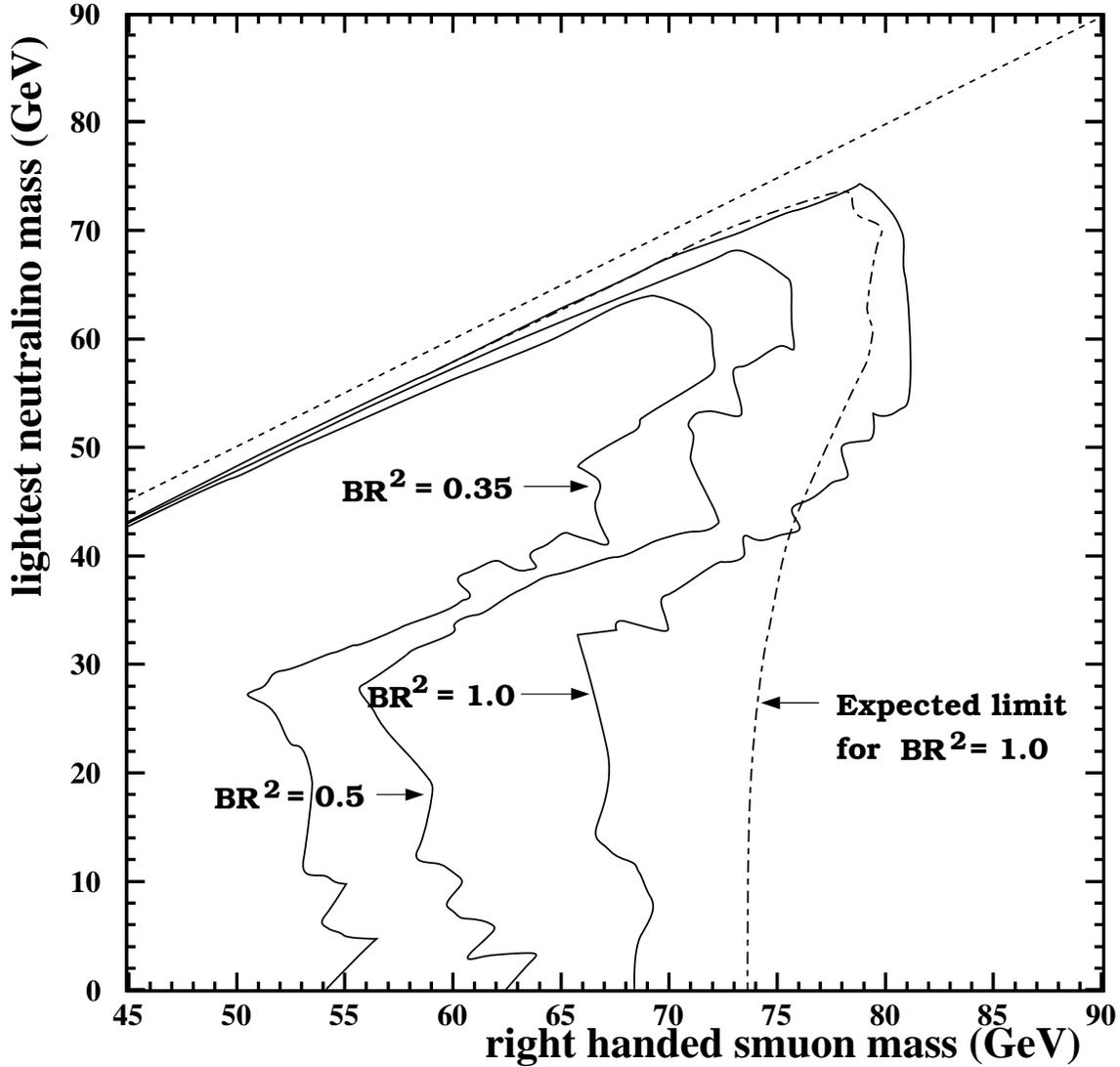}
 \caption{\sl 
95\% CL exclusion region for right-handed smuon pair 
production obtained by combining the $\protect\sqrt{s} = 161-183$~GeV 
data-sets.
The limits are calculated for several values of
the branching ratio squared for 
$\smu^\pm_R \rightarrow  {\mu^\pm} \nt_1$ that are indicated in the figure.
Otherwise they have no supersymmetry model assumptions.
The kinematically allowed region is indicated by the dashed line.  The
expected limit for BR$^2$~=~1.0, calculated from \mc\ alone, is indicated by
the dash-dotted line.
} 
\label{fig-mssm_2}
\end{figure}

\begin{figure}[htbp]
 \epsfxsize=\textwidth 
 \epsffile[0 0 580 600]{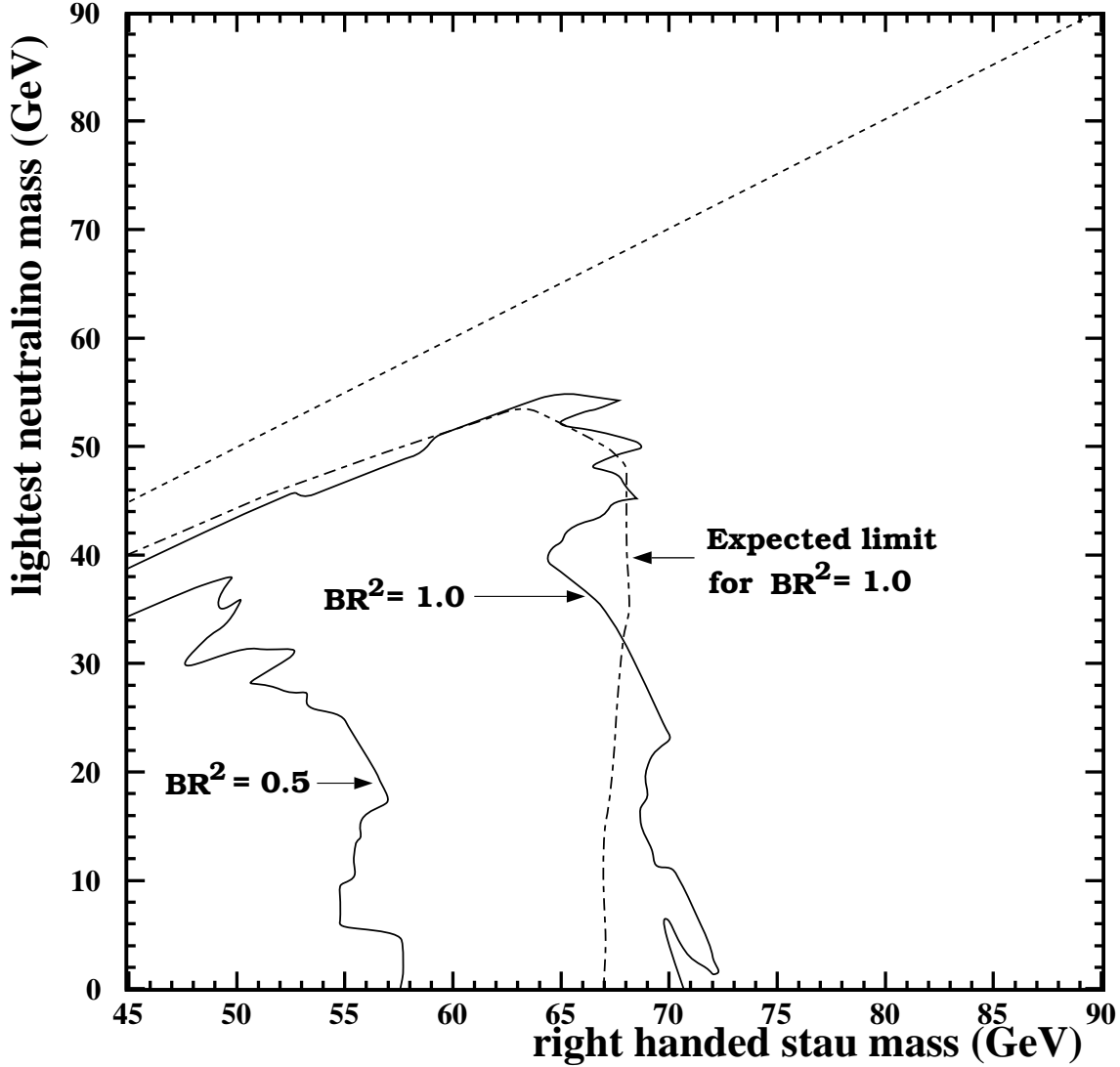}
 \caption{\sl 
95\% CL exclusion region for right-handed stau pair 
production obtained by combining the $\protect\sqrt{s} = 161-183$~GeV 
data-sets.
The limits are calculated for two values of
the branching ratio squared for 
$\stau^\pm_R \rightarrow  {\tau^\pm} \nt_1$.
The selection efficiency for \staupair\ is calculated for the case
that the decay \dstau\ produces unpolarized $\tau^\pm$.
Otherwise the limits have no supersymmetry model assumptions.
The kinematically allowed region is indicated by the dashed
line.  The expected limit for BR$^2$~=~1.0, 
calculated from \mc\ alone, is indicated by
the dash-dotted line.
} 
\label{fig-mssm_3}
\end{figure}

We can use our data to set limits on the masses of right-handed 
sleptons\footnote{
The right-handed slepton is expected to be lighter
than the left-handed slepton. The
right-handed one tends (not generally valid for selectrons)
to
have a lower pair production cross-section, and so
conventionally limits are given for this (usually) conservative case.}
based on the expected right-handed slepton pair cross-sections and
branching ratios.
The cross-sections have been calculated using {\sc Susygen}
at each centre-of-mass energy 
and take into account initial state radiation.
In figure~\ref{fig-mssm_2} we show  limits on right-handed smuons
 as a function of smuon mass and lightest
neutralino mass for several assumed values of
the branching ratio squared for $\smu^\pm_R \rightarrow  {\mu^\pm} \nt_1$.
The expected limit, calculated using \mc\ only, for a branching ratio
of 100\% is also
shown.  Note that the actual limit is lower than expected for large
\dm\ as a result of the excess of events seen for smuons at \seight .
For a branching ratio $\smu^\pm_R \rightarrow  {\mu^\pm} \nt_1$ of
100\% and for a smuon-neutralino mass difference exceeding 2~GeV,
right-handed smuons are excluded at 95\% CL for 
masses below 65~GeV.
The  95\% CL upper limit on the  production of
right-handed \staupair\ times the
 branching ratio squared for $\stau^\pm_R \rightarrow  {\tau^\pm} \nt_1$
is shown in figure~\ref{fig-mssm_3}.  The expected limit for a branching ratio
of 100\% is also shown.
For a branching ratio $\stau^\pm_R \rightarrow  {\tau^\pm} \nt_1$ of
100\% and for a stau-neutralino mass difference exceeding 10~GeV,
right-handed staus are excluded at 95\% CL for 
masses below 64~GeV.
For the case of a massless neutralino (or gravitino) and 100\%
branching ratio, right-handed smuons and staus are excluded at 95\% CL for 
masses below 68~GeV and 70~GeV, respectively.

\begin{figure}[htbp]
 \epsfxsize=\textwidth 
 \epsffile[0 0 580 600]{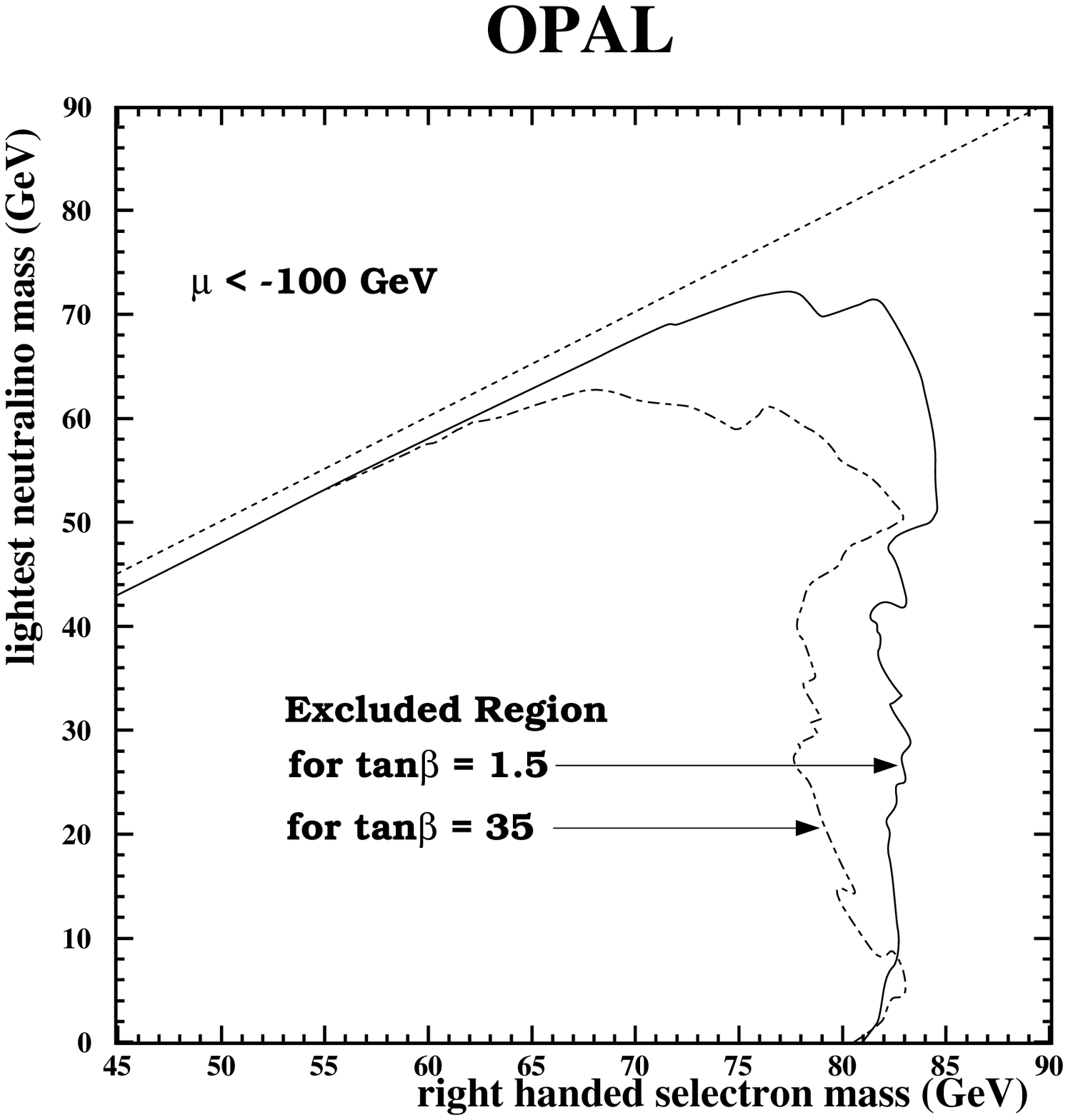}
 \caption{\sl 
For two values of $\tan{\beta}$ and $\mu < -100$~GeV,
95\% CL exclusion regions for right-handed selectron pairs within the MSSM, 
obtained by combining the $\protect\sqrt{s} = 161-183$~GeV data-sets.
The excluded regions are calculated 
taking into account the 
predicted branching ratio for 
$\sele^\pm_R \rightarrow  {\mathrm{e}^\pm} \nt_1$.
The  gauge unification relation,
$M_1 =  \frac{5}{3} \tan^2 \theta_W M_2$, is assumed in calculating the
MSSM cross-sections and branching ratios.
The kinematically allowed region is indicated by the dashed line.
} 
\label{fig-mssm_1}
\end{figure}

\begin{figure}[htbp]
 \epsfxsize=\textwidth 
 \epsffile[0 0 580 600]{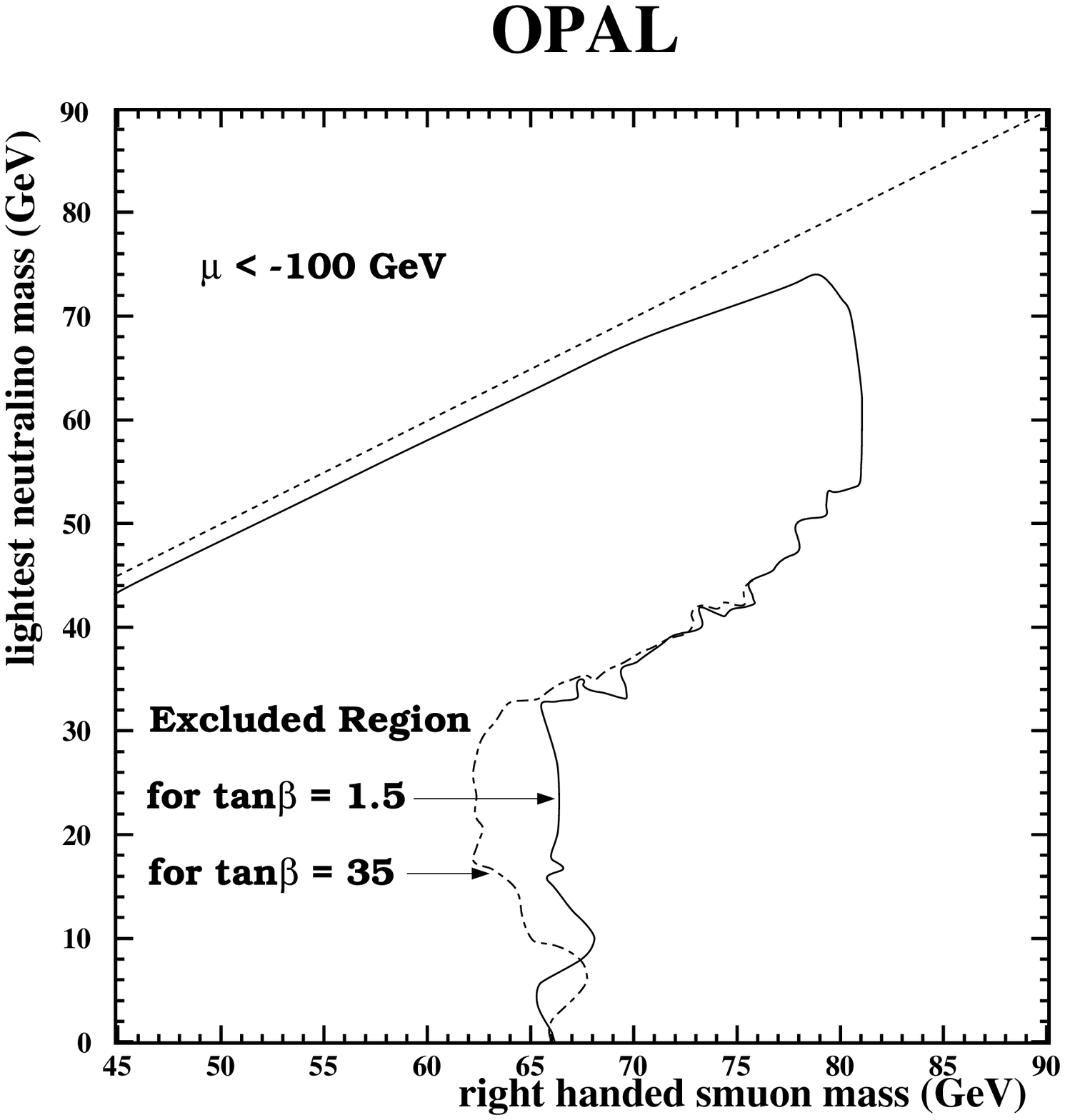}
 \caption{\sl 
For two values of $\tan{\beta}$ and $\mu < -100$~GeV,
 95\% CL exclusion regions for right-handed smuon pairs  within the MSSM,
obtained by combining the $\protect\sqrt{s} = 161-183$~GeV data-sets.
The excluded regions are calculated 
taking into account the 
predicted branchingz ratio for $\smu^\pm_R \rightarrow  {\mu^\pm} \nt_1$.
The  gauge unification relation,
$M_1 =  \frac{5}{3} \tan^2 \theta_W M_2$, is assumed in calculating the
MSSM branching ratios.
The kinematically allowed region is indicated by the dashed line.
} 
\label{fig-mssm_2a}
\end{figure}

\begin{figure}[htbp]
 \epsfxsize=\textwidth 
 \epsffile[0 0 580 600]{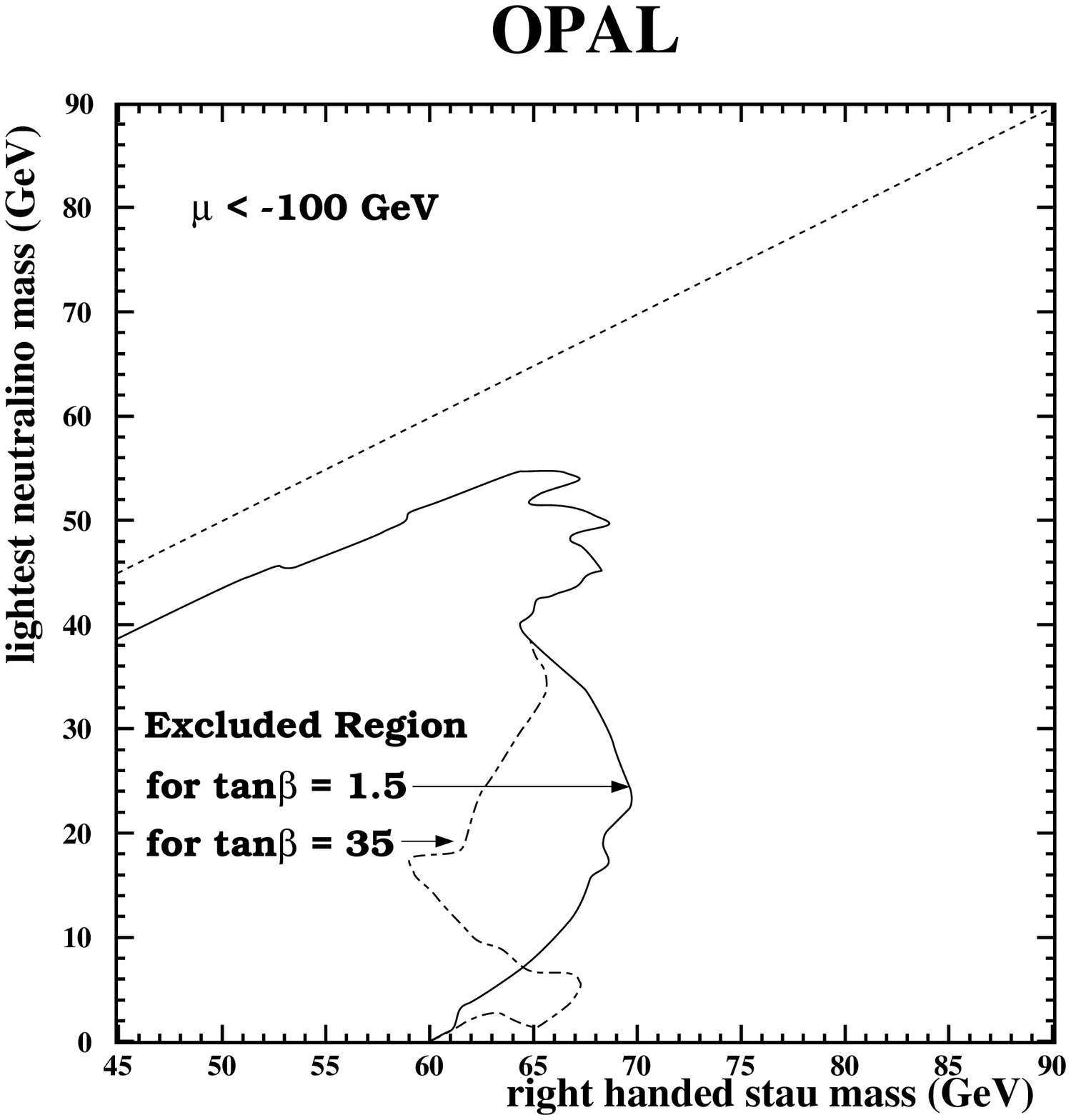}
 \caption{\sl 
For two values of $\tan{\beta}$ and $\mu < -100$~GeV,
 95\% CL exclusion regions for right-handed stau pairs within the MSSM,
obtained by combining the $\protect\sqrt{s} = 161-183$~GeV data-sets.
The excluded regions are calculated 
taking into account the 
predicted branching ratio for $\stau^\pm_R \rightarrow  {\tau^\pm}
\nt_1$.
The  gauge unification relation,
$M_1 =  \frac{5}{3} \tan^2 \theta_W M_2$, is assumed in calculating the
MSSM branching ratios.
The selection efficiency for \staupair\ is calculated for the case
that the decay \dstau\ produces unpolarized $\tau^\pm$.
The kinematically allowed region is indicated by the dashed line.
} 
\label{fig-mssm_3a}
\end{figure}

An alternative approach is to set limits
taking into account the 
predicted cross-section and  branching ratio 
for specific choices of the parameters within the 
Minimal Supersymmetric Standard Model (MSSM)\footnote{
In particular 
 regions of the MSSM parameter space, the branching ratio for 
$\sell^\pm \rightarrow  {\ell^\pm} \nt_1$  
can be essentially zero and so 
 it is not possible to provide general limits on sleptons within the MSSM
 on the basis of this search alone.
The predicted cross-sections and  branching ratios within the MSSM 
are obtained using {\sc Susygen}
and are calculated with the gauge unification relation,
$M_1 =  \frac{5}{3} \tan^2 \theta_W M_2$.}.
For $\mu < -100$~GeV and for two
values of $\tan{\beta}$ (1.5 and 35),
figures~\ref{fig-mssm_1},~\ref{fig-mssm_2a} and~\ref{fig-mssm_3a}
 show 95\% CL exclusion regions 
in the ($m_{\tilde{\ell}^\pm_{\mathrm{R}}}$, $m_{\nt_1}$) 
plane
for right-handed selectrons, smuons and staus, respectively.
For $\mu < -100$~GeV and $\tan{\beta}=1.5$, right-handed sleptons are 
excluded at
95\% CL as follows:
selectrons with masses below 77~GeV for \mbox{$\msele - \mchz > 5$}~GeV;
smuons with masses below 65~GeV for \mbox{$\msmu - \mchz > 2$}~GeV;
and staus with masses below 60~GeV for \mbox{$\mstau - \mchz > 9$}~GeV.

\section{Summary and Conclusions}
 
A selection of di-lepton events with significant missing transverse momentum 
is performed using a total data sample of 77.0~pb$^{-1}$
at e$^+$e$^-$ centre-of-mass energies of 161, 172 and 183~GeV.
The observed numbers of events, four at 161 GeV, nine at 172 GeV and 78
at 183~GeV,
are consistent with the numbers expected from Standard Model processes, 
dominantly arising from \wpair\ production with each W decaying
leptonically.

Further event selection criteria are employed to
search for the pair production of charged scalar leptons, 
leptonically decaying charged Higgs bosons 
and charginos that decay to produce a charged lepton and one or more
invisible particles.
The sensitivity to
new physics is maximised by using an algorithm to optimise the
cut values as functions of the masses of the pair produced new
particle and the unobserved particle produced in its decay.
No evidence for new phenomena is apparent and model independent limits 
on the production cross-section times branching ratio squared
for each new physics process are presented.

Assuming a 100\% branching ratio for the decay
$\sell^\pm_R \rightarrow  {\ell^\pm} \nt_1$, we exclude at 95\% CL:
right-handed smuons with masses below 65~GeV for 
\mbox{$\msmu - \mchz > 2$}~GeV and
right-handed staus with masses below 64~GeV for 
\mbox{$\mstau - \mchz > 10$}~GeV.
Right-handed selectrons are excluded at 95\% CL for 
masses below 77~GeV for \mbox{$\msele - \mchz > 5$}~GeV
within the framework of the
 MSSM assuming
$\mu < -100$~GeV and $\tan{\beta} = 1.5$.

\bigskip\bigskip
\noindent {\Large\bf Acknowledgements.}

\noindent We particularly wish to thank the SL Division for 
the efficient operation
of the LEP accelerator at all energies
 and for their continuing close cooperation with
our experimental group.  We thank our colleagues from CEA, DAPNIA/SPP,
CE-Saclay for their efforts over the years on the time-of-flight and trigger
systems which we continue to use.  In addition to the support staff at our own
institutions we are pleased to acknowledge the  \\
Department of Energy, USA, \\
National Science Foundation, USA, \\
Particle Physics and Astronomy Research Council, UK, \\
Natural Sciences and Engineering Research Council, Canada, \\
Israel Science Foundation, administered by the Israel
Academy of Science and Humanities, \\
Minerva Gesellschaft, \\
Benoziyo Center for High Energy Physics,\\
Japanese Ministry of Education, Science and Culture (the
Monbusho) and a grant under the Monbusho International
Science Research Program,\\
German Israeli Bi-national Science Foundation (GIF), \\
Bundesministerium f\"ur Bildung, Wissenschaft,
Forschung und Technologie, Germany, \\
National Research Council of Canada, \\
Research Corporation, USA,\\
Hungarian Foundation for Scientific Research, OTKA T-016660, 
T023793 and OTKA F-023259.\\


\clearpage


\clearpage

\clearpage

\clearpage

\clearpage

\clearpage

\clearpage

\clearpage

\clearpage

\clearpage

\clearpage

\clearpage

\clearpage

\clearpage

\end{document}